\documentclass[twoside,11pt]{article}

\usepackage{amsmath, amssymb, amsthm}
\usepackage[ruled,vlined]{algorithm2e}
\usepackage{float}
\usepackage{bm}
\usepackage{placeins}
\usepackage{booktabs}
\usepackage{tabularx}

\usepackage{tikz}
\usetikzlibrary{arrows.meta,positioning}
\providecommand{\bm}[1]{\boldsymbol{#1}}
\newcommand{\indep}{\mathop{\perp\!\!\!\!\perp}}

\newcommand{\Cov}{\operatorname{Cov}}

\newcommand{\pthin}{\mathbin{\circ_p}}
\newcommand{\bthin}{\mathbin{\circ_b}}

\usepackage[preprint]{jmlr2e}

\usepackage{lastpage}
\jmlrheading{23}{2026}{1-\pageref{LastPage}}{1/21; Revised 5/22}{9/22}{21-0000}{Penggang Gao, Ming Cai and Hisayuki Hara}

\ShortHeadings{DAG Identification of PT-SEM}{Penggang Gao and Ming Cai and Hisayuki Hara}
\firstpageno{1}

\begin{document}

\title{Causal DAG Identification for Count Data via Poisson Thinning Structural Equation Models}

\author{\name Penggang Gao \email gao.penggang.27p@st.kyoto-u.ac.jp  \\
       \addr Graduate School of Informatics\\
       Kyoto University\\
       Kyoto, 606-8501, Japan
       \AND
       \name Ming Cai \email cai.ming.52d@st.kyoto-u.ac.jp \\
       \addr Graduate School of Informatics\\
       Kyoto University\\
       Kyoto, 606-8501, Japan
       \AND
       \name Hisayuki Hara \email hara.hisayuki.8k@kyoto-u.ac.jp \\
       \addr Institute for Liberal Arts and Sciences\\
       Kyoto University\\
       Kyoto, 606-8501, Japan
       }

\editor{My editor}

\maketitle

\begin{abstract}
Count-valued variables arise in many scientific and applied settings, yet explicit structural models that allow full identification of causal DAGs from observational data remain limited. The Poisson branching structural causal model (PB-SCM) provides a count-valued analogue of linear structural equation models using binomial thinning and independent Poisson exogenous variables, but its causal DAG is generally only partially identifiable.

Building on this framework, we propose the Poisson thinning structural equation model (PT-SEM), which replaces binomial thinning in PB-SCM with Poisson thinning and allows node-wise exogenous distributions from diverse count-distribution families. Under node-wise regularity conditions, we establish identifiability of the causal DAG, the thinning coefficients, and the node-wise exogenous distributions.
The same identification analysis extends to binomial thinning, yielding full identifiability whenever every nonsink has non-Poisson exogenous noise.
We further develop a structure learning algorithm that optimizes, via dynamic programming, a BIC score based on local likelihoods evaluated at plug-in moment estimates, and establish its consistency for DAG selection.

Simulations demonstrate favorable performance in DAG recovery and thinning-coefficient estimation, and a real-data application illustrates the practical utility of PT-SEM.
\end{abstract}

\begin{keywords}
Bayesian information criterion, cumulant generating functions, dynamic programming, moment estimation, thinning operators
\end{keywords}

\section{Introduction}

Count-valued variables arise across scientific and applied domains, from disease and accident counts to transactions and events recorded in economic and operational systems, and learning their causal structure from observational data is a recurring problem.
In the absence of latent confounding, several functional causal models for continuous variables identify causal directed acyclic graphs (DAGs) under suitable model-specific assumptions.
Representative examples include the linear non-Gaussian acyclic model (LiNGAM)~\citep{shimizu2006linear}, nonlinear additive-noise models (ANMs)~\citep{hoyer2008nonlinear,peters2014causal}, and the post-nonlinear (PNL) causal model~\citep{zhang2009identifiability}.
In contrast, conventional causal graph learning methods such as the Peter-Clark (PC) algorithm~\citep{spirtes2001causation} and greedy equivalence search (GES)~\citep{chickering2002optimal} generally recover only a Markov equivalence class under the Markov and faithfulness assumptions.
For count data, however, explicit SEMs with full DAG identifiability from observational data remain limited.

A major line of work develops identifiable DAG models for count data by directly specifying the distribution of each node given its parents.
\citet{park2015learning} introduced Poisson DAG models together with an overdispersion-based score; subsequent work extended this line to quadratic variance functions and generalized hypergeometric DAG models~\citep{park2018learning,park2019identifiability}.
Zero-inflated extensions include the Poisson Bayesian network~\citep{choi2020bayesian} and the more general ZiG-DAG model~\citep{choi2023model}.
At a broader level, conditionally parametric causal models establish identifiability for several parametric families, including Poisson~\citep{bodik2025identifiability}.
Across these models, identifiability is derived from restrictions on the node-wise conditional distributions.

A complementary line of work constructs explicit count-valued SEMs using thinning operators.
\citet{qiao2024causal} introduced the Poisson branching structural causal model (PB-SCM), which combines binomial thinning~\citep{Steutel1979} with independent Poisson exogenous noise, and developed higher-order cumulant criteria for identifying causal directions under graphical conditions.
Building on this model, \citet{NEURIPS2024_15aaa922} derived its probability generating function and established conditions for recovering the skeleton and identifiable local orientations.
More recently, \citet{qiao2026latent} extended PB-SCM to latent confounding and studied the identifiability of local three-variable structures.
Despite these developments, the full DAG of the causally sufficient PB-SCM is not identifiable in general.
Moreover, its edge coefficients are restricted to \((0,1]\), and all exogenous variables are Poisson.

We address this gap by introducing the Poisson thinning structural equation model (PT-SEM), built on a generalized thinning operator with Poisson offspring, 
in which each unit of an integer-valued input independently produces a Poisson-distributed offspring count 
\citep{latour1998existence}. 
Whereas the Bernoulli offspring in binomial thinning represent survival or retention, Poisson offspring permit one-to-many reproduction, with the thinning parameter interpreted as the mean offspring count rather than a probability \citep{chen2022cluster,yu2023obsinar}. 
In PT-SEM, each thinning coefficient enters linearly in the conditional mean and can take any value in \([0,\infty)\), while the mutually independent exogenous variables need not follow a common distributional family.

Despite this distributional freedom, under node-wise regularity conditions, the observational distribution uniquely determines the DAG, the thinning coefficient matrix, and the exogenous distributions. The proof rests on a characterization of sink nodes through conditional cumulant generating functions. 
The same analysis shows that the Poisson family is exceptional under binomial thinning: if every exogenous distribution is non-Poisson, the DAG and the binomial thinning coefficients are identifiable.

For structure learning, we construct a decomposable plug-in BIC score by combining moment-based local parameter estimates with a finite family of candidate exogenous distributions, thereby avoiding repeated numerical likelihood maximization during structure search. The score is optimized exactly by subset dynamic programming \citep{singh2005finding,SilanderMyllymaki2006UAI}.
We further establish consistency of the resulting plug-in BIC for DAG structure and exogenous-family selection.

The remainder of the paper is organized as follows.
Sections~\ref{sec:model} and~\ref{sec:identifiability} introduce PT-SEM and develop its identification theory. Section~\ref{sec:learning} presents the structure-learning procedure and its asymptotic guarantees. Section~\ref{sec:experiments} reports the numerical experiments, and Section~\ref{sec:real-world}  reports an application to real-world event-count analysis. Section~\ref{sec:conclusion}  concludes the paper.
All proofs of theoretical results stated in the main text can be found in Appendix~\ref{sec:appendix}.
Code and reproducibility materials for the numerical experiments and real-data analysis are available at \url{https://github.com/KOUHOUKOU/PT-SEM}.
\section{Poisson Thinning Structural Equation Model}
\label{sec:model}

\subsection{Basic facts on SEMs and Poisson Thinning}
\label{subsec:poisson-thinning}
Let \(\bm X=(X_1,\ldots,X_d)^\top\) be a random vector taking values in \(\mathbb N_0^d\), where \(\mathbb N_0=\{0,1,2,\ldots\}\), and let \(V=[d]\). Let \(G=(V,E)\) be a directed acyclic graph (DAG). We write \(\operatorname{pa}_G(i)\) and \(\operatorname{ch}_G(i)\) for the parent and child sets of $i \in V$, 
respectively. 
For \(U\subseteq V\), write \(G[U]\) for the subgraph induced by \(U\), with edge set \(E[U]=E\cap(U\times U)\). A node \(i\in U\) is a sink in \(G[U]\) if it has no children in \(U\), that is, \(\operatorname{ch}_G(i)\cap U=\varnothing\). 
For \(S\subseteq V\), 
define \(\bm X_S=(X_j)_{j\in S}\).

A SEM defined by a DAG \(G\) induces a distribution that is Markov with respect to \(G\), 
\[
p_{\bm X}(\bm x)=\prod_{i\in V}p_{X_i\mid\bm X_{\operatorname{pa}_G(i)}}\!\left(x_i\mid\bm x_{\operatorname{pa}_G(i)}\right),
\quad
\bm x\in\mathbb N_0^d.
\]

Thinning operators provide a discrete analogue of multiplication for nonnegative integer-valued variables. The classical binomial thinning operator, introduced by \cite{Steutel1979}, is defined by
\[
\alpha \circ_b X = \sum_{i=1}^{X} B_i, \quad B_i \stackrel{\mathrm{iid}}{\sim} \mathrm{Bernoulli}(\alpha), 
\]
where Bernoulli offspring variables \( B_1,B_2,\ldots \) are independent of \(X\). This operator has been adopted as the basic propagation mechanism in the INAR model of \citet{alosh1987inar}, and subsequent thinning-based developments are reviewed by \citet{weiss2008thinning} and \citet{Scotto2015}. 
Using the binomial thinning operator, \citet{qiao2024causal} proposed the Poisson branching structural causal model (PB-SCM), defined as follows. 
\begin{definition}[\citealp{qiao2024causal}]
\label{def:pbscm}
A Poisson branching structural causal model (PB-SCM) associated with $G$ is defined by 
\begin{equation}
\label{model:pbscm}
X_i = \sum_{j \in \operatorname{pa}_G(i)} \alpha_{ij} \circ_b X_j + \varepsilon_i, \quad i \in V, 
\end{equation}
where \(\alpha_{ij} \in (0,1]\) for \(j\to i\in E\). 
All Bernoulli offspring variables associated with $\alpha_{ij}\circ_b X_j$ are mutually independent and independent of $\{\varepsilon_i : i\in V\}$. 
\(\varepsilon_1,\ldots,\varepsilon_d\) are mutually independent Poisson variables.
\end{definition}
It is straightforward to verify that
\begin{equation}
\label{eq:ptsem-conditional-mean}
\mathbb{E}\!\left[X_i\mid\bm X_{\operatorname{pa}_G(i)}\right]
=
\mathbb{E}[\varepsilon_i]
+
\sum_{j\in\operatorname{pa}_G(i)}\alpha_{ij}X_j.
\end{equation}
Hence, the thinning coefficient $\alpha_{ij}$ acts as the direct effect of $X_j$ on the conditional mean of $X_i$. However, since the binomial thinning operator requires $\alpha_{ij}\in(0,1]$, the expressive power of the resulting structural equation model is inherently limited. Moreover, PB-SCM also restricts the exogenous noise variables to follow Poisson distributions and provides only partial identifiability of the underlying DAG. 

To address these limitations, we generalize the thinning mechanism while retaining the interpretation of the thinning coefficients as direct effects on the conditional mean.
Generalized thinning extends binomial thinning by replacing Bernoulli offspring with more general nonnegative integer-valued offspring variables \citep{latour1998existence}. Among these generalized thinning operators, we consider the case where the offspring distribution is Poisson, referred to as Poisson thinning \citep{chen2022cluster,yu2023obsinar}. 

\begin{definition}[e.g., \citealp{chen2022cluster,yu2023obsinar}]
Let \(X\) be a nonnegative integer-valued random variable and let \(\alpha\ge0\). The Poisson thinning operator is defined by
\[
\alpha\circ_p X=\sum_{\ell=1}^{X}Z_\ell,\quad Z_\ell\overset{\mathrm{iid}}{\sim}\mathrm{Poisson}(\alpha),
\]
where Poisson offspring variables  \(Z_1,Z_2,\ldots\)  are independent of \(X\). The empty sum is zero. 
\end{definition}

For later use, we record the elementary moment identities. Suppose the Poisson offspring defining \(\alpha\circ_p X\) are independent of \(Y\) below; then, whenever the relevant moments exist,
\begin{equation}
\label{eq:ptsem-moment}
\begin{gathered}
    \mathbb{E}[\alpha\circ_p X]
    =
    \alpha\mathbb{E}[X],
    \quad
    \operatorname{Var}(\alpha\circ_p X)
    =
    \alpha^2\operatorname{Var}(X)
    +
    \alpha\mathbb{E}[X],\\
    \operatorname{Cov}(\alpha\circ_p X,Y)
    =   
    \alpha\operatorname{Cov}(X,Y).    
\end{gathered}
\end{equation}
The moment generating function (MGF) of $\alpha \circ_p X$ satisfies
\begin{equation}
    \label{eq:poisson-thinning-mgf}
    M_{\alpha\circ_p X}(t)=M_X\left(\alpha(e^t-1)\right),
\end{equation}
at values of \(t\) for which the expressions are finite. 

\subsection{Poisson thinning structural equation model}
\label{subsec:ptsem-definition}
We define the proposed Poisson thinning structural equation model as follows.
\begin{definition}[Poisson thinning structural equation model]
\label{def:ptsem}
A Poisson thinning structural equation model (PT-SEM) associated with \(G\) is defined by
\begin{equation}
    \label{model:PT-SEM}
    X_i=\sum_{j\in\operatorname{pa}_G(i)}\alpha_{ij}\circ_p X_j+\varepsilon_i,\quad i\in V,
\end{equation}
where \(\alpha_{ij}>0\) for \(j\to i\in E\). 
All Poisson offspring variables associated with $\alpha_{ij}\circ_p X_j$ are mutually independent and independent of $\{\varepsilon_i : i\in V\}$.
\(\varepsilon_1,\ldots,\varepsilon_d\) are mutually independent count-valued random variables. No common parametric family is imposed on these exogenous distributions at the model-definition level.
\end{definition}

We use the minimal graph convention throughout: a directed edge \(j\to i\) is present if and only if \(\alpha_{ij}>0\), and we set \(\alpha_{ij}=0\) whenever \(j\notin\operatorname{pa}_G(i)\). Thus the DAG \(G\) is the support graph of the thinning coefficients in \eqref{model:PT-SEM}.
PT-SEM \eqref{model:PT-SEM} also satisfies \eqref{eq:ptsem-conditional-mean}. 
Hence, the thinning coefficient \(\alpha_{ij}\) is interpreted as a direct effect of \(X_j\) on the conditional mean of \(X_i\). Unlike PB-SCM, PT-SEM does not restrict thinning coefficients to the unit interval.

For notational convenience, we also write \eqref{model:PT-SEM} in compact matrix form
\begin{equation}
    \label{model:PT-SEM-vec}
    \bm X=\bm A\circ_p\bm X+\bm\varepsilon,
\end{equation}
where \(\bm\varepsilon=(\varepsilon_1,\ldots,\varepsilon_d)^\top\), \(\bm A=(\alpha_{ij})\in[0,\infty)^{d\times d}\), and the positive entries of \(\bm A\) correspond exactly to the directed edges of \(G\). Since \(G\) is acyclic, the equations define \(\bm X\) recursively along any topological order; equivalently, after a topological ordering of the variables, the coefficient matrix can be written as a strictly lower triangular matrix. This recursive construction yields the following Markov property.

\begin{proposition}
\label{prop:ptsem-markov}
The observational distribution induced by the PT-SEM in Definition~\ref{def:ptsem} is Markov with respect to \(G\). Equivalently, its probability mass function factorizes as
\[
p_{\bm X}(\bm x)=\prod_{i\in V}
p_{X_i\mid\bm X_{\operatorname{pa}_G(i)}}\!\left(x_i\mid\bm x_{\operatorname{pa}_G(i)}\right),
\quad
\bm x\in\mathbb N_0^d.
\]
\end{proposition}

PT-SEM also connects to conditional-distribution models for count DAGs. 
If all exogenous variables are Poisson, with \(\varepsilon_i\sim\operatorname{Poisson}(\lambda_i)\) for every \(i\in V\), then we have
\[
X_i\mid\bm X_{\operatorname{pa}_G(i)}\sim\operatorname{Poisson}\!\left(\lambda_i+\sum_{j\in\operatorname{pa}_G(i)}\alpha_{ij}X_j\right),
\quad
i\in V.
\]
Thus the resulting PT-SEM is a Poisson DAG \citep{park2015learning} with identity link.

\section{Identifiability of PT-SEM}
\label{sec:identifiability}
In this section, we establish the identifiability of PT-SEM under mild regularity conditions on the exogenous distributions. 
According to Definition \ref{def:ptsem}, PT-SEM does not impose a common distributional family on the exogenous noises. In the following, assume that the exogenous variables \(\varepsilon_1,\ldots,\varepsilon_d\) are mutually independent and satisfy the following conditions:
\begin{description}
\item[\textup{(A1)}]\phantomsection\label{ass:zero-mass}
\(\Pr(\varepsilon_i=0)>0\) for every \(i\in V\).
\item[\textup{(A2)}]\phantomsection\label{ass:nondegenerate-noise}
\(\operatorname{Var}(\varepsilon_i)>0\) for every \(i\in V\).
\item[\textup{(A3)}]\phantomsection\label{ass:local-mgf}
\(M_{\varepsilon_i}(t)<\infty\) in a neighborhood of zero for every \(i\in V\).
\end{description}

These conditions hold for common nondegenerate count distributions such as Poisson, zero-inflated Poisson, negative binomial, and geometric families. 
Proofs of all lemmas, propositions, and theorems in this section are given in Appendix~\ref{app:proofs-section3}.

\subsection{Identifiability in the Bivariate Case}
\label{subsec:bivariate-mechanism}

We first establish the identifiability of the causal direction in the bivariate PT-SEM
\begin{equation}
\label{eq:bivariate-ptsem}
X_1=\varepsilon_1,\quad
X_2=\alpha\pthin X_1+\varepsilon_2,\quad
\alpha>0,
\end{equation}
where \(\varepsilon_1\) and \(\varepsilon_2\) are independent exogenous variables satisfying \hyperref[ass:zero-mass]{\textup{(A1)}}--\hyperref[ass:local-mgf]{\textup{(A3)}}. 
To prove identifiability, it suffices to rule out the opposite causal direction. Suppose, for contradiction, that the observational distribution generated by \eqref{eq:bivariate-ptsem} admits a reverse PT-SEM representation
\begin{equation}
\label{eq:reverse-ptsem}
X_2=\varepsilon_2^*,\quad
X_1=\beta\pthin X_2+\varepsilon_1^*,\quad
\beta\ge0,
\end{equation}
where \(\varepsilon_1^*\) is independent of \(X_2\). We show that this assumption leads to a contradiction, and therefore no reverse PT-SEM can generate the same observational distribution.

For \(s=0,1,2\), define the conditional MGFs
\[
M_s(t)
=
\mathbb{E}\!\left[\exp(tX_1)\mid X_2=s\right].
\]
These functions are completely determined by the joint distribution of \((X_1,X_2)\). Consequently, both the forward and hypothetical reverse PT-SEM representations must induce the same functions \(M_0,M_1,\) and \(M_2\).

The proof proceeds in three steps. We first derive structural identities satisfied by the conditional MGFs under the forward PT-SEM \eqref{eq:bivariate-ptsem}. We then derive a different identity that any reverse PT-SEM representation must satisfy. Finally, we show that these two sets of identities are algebraically incompatible, thereby ruling out the reverse representation. 

\begin{lemma}
\label{lem:bivariate-three-layer-bayes-tilt}
Under the forward PT-SEM \eqref{eq:bivariate-ptsem}, the conditional distribution defining \(M_0\) is nondegenerate, and there are constants satisfying \(\gamma_0,\delta_0,\delta_1\ge0\) and \(\gamma_1,\delta_2>0\) such that, for all \(t\) near zero,
\begin{equation}
    \label{eq:bivariate-three-layer-tilt}
    \begin{aligned}
        M_1(t)
        &=
        \gamma_0M_0(t)
        +
        \gamma_1M_0'(t), \\
        M_2(t)  
        &=
        \delta_0M_0(t)
        +
        \delta_1M_0'(t)
        +
        \delta_2M_0''(t).
    \end{aligned}
\end{equation}
\end{lemma}


Lemma~\ref{lem:bivariate-three-layer-bayes-tilt} characterizes the conditional MGFs under the forward PT-SEM \eqref{eq:bivariate-ptsem}. Its proof uses Bayes' formula together with the Poisson--exogenous convolution structure to express \(M_1\) and \(M_2\) as linear combinations of \(M_0\) and its derivatives.

\begin{lemma}
\label{lem:reverse-ptsem-layer-identity}
Under the above reverse PT-SEM~\eqref{eq:reverse-ptsem}, 
\begin{equation}
    \label{eq:false-poisson-sink-tilt}
    M_s(t) = M_0(t)\exp\{s\beta(e^t-1)\},
    \quad s=0,1,2.
\end{equation}
\end{lemma}

Lemma~\ref{lem:reverse-ptsem-layer-identity} gives the characterization under a hypothetical reverse PT-SEM. Its proof follows directly from the conditional distribution of the Poisson thinning term.

\begin{lemma}
\label{lem:three-layer-incompatibility}

Suppose that \(M_0,M_1,\) and \(M_2\) satisfy \eqref{eq:bivariate-three-layer-tilt}. Then there is no \(\beta\ge0\) such that \eqref{eq:false-poisson-sink-tilt} holds for all \(t\) in a neighborhood of zero.
\end{lemma}

Lemma~\ref{lem:three-layer-incompatibility} shows that the identities \eqref{eq:bivariate-three-layer-tilt} and \eqref{eq:false-poisson-sink-tilt} cannot hold simultaneously. Therefore, the forward and reverse PT-SEMs impose incompatible constraints on the same observational conditional MGFs, yielding the following result.

\begin{proposition}
\label{prop:bivariate-directional-asymmetry}
Under \hyperref[ass:zero-mass]{\textup{(A1)}}--\hyperref[ass:local-mgf]{\textup{(A3)}}, the bivariate PT-SEM \eqref{eq:bivariate-ptsem} admits no reverse PT-SEM representation \(X_2\to X_1\). Hence, \(X_1\to X_2\) is identifiable from the joint distribution.
\end{proposition}


The multivariate argument below localizes the same contradiction around a nonsink node and one of its children, after fixing the remaining variables at their zero configuration.
\subsection{Local sink characterization}
\label{subsec:local-sink-characterization}

To establish identifiability for general multivariate PT-SEMs, we first show how sink nodes can be identified from the observational distribution by extending the bivariate argument. 
Let \(U \subseteq V\) satisfy \(\operatorname{pa}_G(u)\subseteq U\) for every \(u\in U\). 
Such a subset is called parent-closed.
This condition ensures that \(\bm X_U\) follows the PT-SEM associated with the induced subDAG \(G[U]\).
For \(i\in U\), define a conditional cumulant generating function (CGF)
\[
K_i^U(t\mid\bm x)
=
\log
\mathbb{E}\!\left[
\exp(tX_i)
\mid
\bm X_{U\setminus\{i\}}=\bm x
\right],
\]
for 
\(\bm x\) with positive probability.

If \(i\) is not a sink in \(G[U]\), choose a child \(c\in\operatorname{ch}_{G[U]}(i)\), set \(R=U\setminus\{i,c\}\), and let \(\bm 0_R\) denote the zero vector indexed by \(R\). For \(s=0,1,2\), define the local conditioning events
\[
L_s=\{X_c=s,\ \bm X_R=\bm 0_R\},
\]
with the convention that \(L_s=\{X_c=s\}\) when \(R=\varnothing\).
The conditional MGFs
\begin{equation}
\label{eq:cond_mgf_sink}
M_s(t)
=
\mathbb{E}\!\left[\exp(tX_i)\mid L_s\right],
\quad
s=0,1,2,
\end{equation}
serve as local analogues of
\(\mathbb{E}[\exp(tX_1)\mid X_2=s]\)
in the bivariate analysis.
The following lemma extends Lemma~\ref{lem:bivariate-three-layer-bayes-tilt} from the bivariate case to general PT-SEMs. 
\begin{lemma}
\label{lem:local-three-layer-bayes-tilt}
Let \(U\) be parent-closed, and suppose that \(i\in U\) is not a sink in \(G[U]\). With \(c\), \(R\), \(L_s\), and \(M_s\) defined as above, the conditional distribution defining \(M_0\) is nondegenerate, and there exist constants \(\gamma_0,\delta_0,\delta_1\ge0\) and \(\gamma_1,\delta_2>0\) such that, for all \(t\) near zero,
\begin{equation}
\label{eq:local-three-layer-identities}
\begin{aligned}
M_1(t)&=\gamma_0M_0(t)+\gamma_1M_0'(t),\\
M_2(t)&=\delta_0M_0(t)+\delta_1M_0'(t)+\delta_2M_0''(t).
\end{aligned}
\end{equation}
\end{lemma}
Lemma~\ref{lem:local-three-layer-bayes-tilt} leads to the following characterization for a node to be a sink. 

\begin{proposition}
\label{prop:conditional-cgf-sink}
Let \(U\) be parent-closed and let \(i\in U\). Then \(i\) is a sink in \(G[U]\) if and only if there exists a CGF \(g\) of a nondegenerate count-valued distribution and coefficients \(b_j\ge0\), \(j\in U\setminus\{i\}\), such that
\begin{equation}
\label{eq:sink-conditional-cgf}
K_i^U(t\mid\bm x)=g(t)+(e^t-1)\sum_{j\in U\setminus\{i\}}b_jx_j
\end{equation}
for all positive-probability conditioning values \(\bm x\) and all \(t\) near zero.
\end{proposition}

Proposition~\ref{prop:conditional-cgf-sink} identifies all sink nodes of $G$ from the observational distribution. Since removing sink nodes leaves a parent-closed induced subDAG, Proposition~\ref{prop:conditional-cgf-sink} can be applied recursively to the remaining DAG to identify its sink nodes. Repeating this procedure yields a topological ordering, which is used in the exact-identifiability theorem below.

\subsection{Exact identifiability of PT-SEM}
\label{subsec:exact-identifiability-ptsem}
Although a recovered topological order is not necessarily unique, any such order is sufficient for identifying the DAG. The remaining task is to determine, for each node, which preceding nodes in the recovered topological order are its true parents. The main idea is as follows.

Fix a recovered topological order, and relabel the variables so that \(X_1,\ldots,X_d\) are indexed according to this order. Then PT-SEM~\eqref{model:PT-SEM} can be written as
\[
X_k
=
\sum_{j<k}\alpha_{kj}\pthin X_j+\varepsilon_k,
\qquad
k=1,\ldots,d,
\]
where absent direct effects have coefficient zero. Let \(S_k=\{1,\ldots,k-1\}\). For \(S_k\neq\varnothing\) define
\[
\bm\Sigma_{S_kS_k}
=
\{\operatorname{Cov}(X_j,X_\ell)\}_{j,\ell\in S_k},
\quad
\bm\Sigma_{S_kk}
=
\{\operatorname{Cov}(X_j,X_k)\}_{j\in S_k},
\quad
\bm\alpha_{k,S_k}
=
(\alpha_{kj})_{j\in S_k}.
\]

\begin{lemma}
\label{lemma:cov_positive_definite}
Under conditions \hyperref[ass:nondegenerate-noise]{\textup{(A2)}} and \hyperref[ass:local-mgf]{\textup{(A3)}}, for any nonempty subset \(S\subseteq V\), the covariance submatrix \(\boldsymbol{\Sigma}_{SS}=\{\operatorname{Cov}(X_i,X_j)\}_{i,j\in S}\) is positive definite.
\end{lemma}

The covariance identity for Poisson thinning in \eqref{eq:ptsem-moment}, together with the independence of \(\varepsilon_k\) from \(X_{S_k}\), yields $\bm\Sigma_{S_kk}=\bm\Sigma_{S_kS_k}\bm\alpha_{k,S_k}$.
Since \(\bm\Sigma_{S_kS_k}\) is positive definite by Lemma~\ref{lemma:cov_positive_definite}, \(\bm\alpha_{k,S_k}\) is uniquely determined by the observational distribution. Its positive entries identify the parents of \(X_k\), while its zero entries indicate the absence of the corresponding edges. Hence, the parent set of every node, and therefore the DAG, is uniquely determined by the observational distribution.

Once the DAG and the thinning coefficients have been identified, the node-wise exogenous distributions are recovered as follows. Indeed, the event $\{\bm X_{S_k}=\bm0_{S_k}\}$ has positive probability by \hyperref[ass:zero-mass]{\textup{(A1)}} and the mutual independence of the exogenous variables. On this event, all incoming thinning terms into \(X_k\) vanish, so
\[
\Pr(\varepsilon_k=r)=\Pr\!\left(X_k=r\mid\bm X_{S_k}=\bm0_{S_k}\right),\quad
r\in\mathbb N_0,
\]
with the usual convention when \(S_k=\varnothing\). Thus, the node-wise exogenous distributions are also uniquely determined by the observational distribution. These observations lead to the following theorem, whose complete proof is given in Appendix~\ref{app:proofs-section3}. 
\begin{theorem}
\label{thm:ptsem-exact-identifiability}
Consider a PT-SEM on $G$ satisfying \hyperref[ass:zero-mass]{\textup{(A1)}}--\hyperref[ass:local-mgf]{\textup{(A3)}}. Then the observational distribution of \(\bm X\) uniquely determines \(G\), the thinning coefficient matrix \(\bm A=(\alpha_{ij})_{i,j\in V}\), and the node-wise exogenous distributions.
\end{theorem}
\subsection{Identifiability of PB-SCM}
\label{subsec:extension-binomial-thinning}
As shown in \cite{qiao2024causal}, the PB-SCM in Definition~\ref{def:pbscm} is not fully identifiable. This non-identifiability stems from the assumption of Poisson exogenous noise. We show below that the PB-SCM \eqref{model:pbscm} becomes identifiable when this restriction is relaxed by allowing the exogenous noise of every nonsink variable to follow a non-Poisson count distribution. The proof follows an argument similar to that used for the identifiability of the PT-SEM, based on a characterization of sink nodes through their conditional cumulant generating functions.

We assume that the exogenous count variables $\varepsilon_1,\ldots,\varepsilon_d$ in \eqref{model:pbscm} are mutually independent and satisfy \hyperref[ass:zero-mass]{\textup{(A1)}}--\hyperref[ass:local-mgf]{\textup{(A3)}}, with the distribution families allowed to differ across nodes.

\begin{proposition}
\label{prop:binomial-poisson-obstruction}
Let \(U\subseteq V\) be parent-closed and let \(i\in U\). Then the following hold.
\begin{enumerate}
    \item If \(i\) is a sink in \(G[U]\), then there exist a cumulant
    generating function \(g\) of a nondegenerate count-valued distribution and coefficients \(b_j\in[0,1]\),
    \(j\in U\setminus\{i\}\), such that
    \begin{equation}
    \label{eq:binomial-sink-cgf}
    K_i^U(t\mid\bm x)
    =
    g(t)
    +
    \sum_{j\in U\setminus\{i\}}
    x_j\log(1-b_j+b_j e^t)
    \end{equation}
    for every positive-probability conditioning value \(\bm x\) and
    all \(t\) near zero.

    \item If \(i\) is a nonsink in \(G[U]\) and
    \eqref{eq:binomial-sink-cgf} holds for some cumulant generating
    function \(g\) of a nondegenerate count-valued distribution and coefficients \(b_j\in[0,1]\),
    \(j\in U\setminus\{i\}\), then \(\varepsilon_i\) must be Poisson.
\end{enumerate}
\end{proposition}

Proposition~\ref{prop:binomial-poisson-obstruction} shows that, when the exogenous noise of every nonsink is non-Poisson, \eqref{eq:binomial-sink-cgf} characterizes the sink nodes. Hence, the sink nodes can be identified from the observational distribution. By recursively identifying and removing sinks, as in the proof of Theorem~\ref{thm:ptsem-exact-identifiability}, we obtain the following identifiability result. 

\begin{theorem}
\label{thm:binomial-nonpoisson-identifiability}
If \(\varepsilon_i\) is non-Poisson for every nonsink \(i\) in \(G\), then the observational distribution of
\eqref{model:pbscm} uniquely determines \(G\), the thinning coefficient matrix \(\bm A=(\alpha_{ij})_{i,j\in V}\), and the node-wise exogenous distributions. 
\end{theorem}
\section{Learning the causal structure in PT-SEM}
\label{sec:learning}

Section~\ref{sec:identifiability} establishes that, without specifying parametric families for the exogenous distributions, the observational distribution of a PT-SEM identifies its DAG, thinning coefficients, and node-wise exogenous distributions. We now turn to the finite-sample problem of selecting a DAG and an exogenous family for each node from a finite set of candidate families, and estimating the associated thinning coefficients. 

Direct likelihood-based search would require repeated numerical optimization of convolutional local likelihoods over many candidate parent sets and families. We instead use parameter estimates based on the method of moments (MoM) and evaluate the local likelihoods at these estimates. Because both the likelihood and the number of free parameters decompose over nodes, the resulting plug-in Bayesian information criterion (BIC; \citealp{schwarz1978estimating}) is decomposable and can be optimized exactly by dynamic programming.

Throughout this section, let \(\mathcal D_N=\{\bm x^{(i)}\}_{i=1}^N\) be an i.i.d. sample from the observational distribution \(P_0\) of a PT-SEM with true DAG \(G_0=(V,E_0)\), where \(V=\{1,\ldots,d\}\), and let \(\mathcal G\) denote the set of all DAGs on \(V\). Let \(\mathcal M=\{\mathcal F_r:r\in\mathcal I\}\) be a finite collection of candidate exogenous families, where \(\mathcal F_r=\{p_r(\cdot;\bm\theta):\bm\theta\in\Theta_r\}\) is a parametric family of probability mass functions on \(\mathbb N_0\). We assume that, for every node \(k\), the true distribution of \(\varepsilon_k\) is represented by at least one family in \(\mathcal M\). We take each \(\Theta_r\) to contain only parameter values for which \(p_r(\cdot;\bm\theta)\) satisfies \hyperref[ass:zero-mass]{\textup{(A1)}}--\hyperref[ass:local-mgf]{\textup{(A3)}}.
\subsection{Likelihood under a fixed DAG and exogenous family assignment}
\label{likelihood}

Fix a candidate model \(m=(G,\bm\tau)\), where \(G=(V,E)\in\mathcal G\) and \(\bm\tau=(\tau_k)_{k\in V}\in\mathcal I^V\) is a node-wise exogenous-family assignment. The assignment \(\bm\tau\) is fixed when evaluating \(m\) and will be selected using BIC in Section~\ref{subsec:bic}.
For scoring, thinning coefficients are allowed to be zero, so candidate DAGs may contain redundant edges. 

For a fixed node \(k\in V\), set \(S=\operatorname{pa}_G(k)\) and \(r=\tau_k\). Let \(\bm\alpha_{k,S}=(\alpha_{kj})_{j\in S}\), and let \(\bm\theta_k\in\Theta_r\) denote the parameter of \(\mathcal F_r\) at node \(k\). For \(\bm x_S=(x_j)_{j\in S}\), the model implies that, conditional on \(\bm X_S=\bm x_S\), the incoming thinning terms sum to a Poisson random variable with mean \(\bm\alpha_{k,S}^{\top}\bm x_S\), where the empty sum is zero. Hence, the induced conditional pmf is
\[
p_{k,r}\!\left(x_k\mid\bm x_S;\bm\alpha_{k,S},\bm\theta_k\right)=\sum_{t=0}^{x_k}e^{-\bm\alpha_{k,S}^{\top}\bm x_S}\frac{\left(\bm\alpha_{k,S}^{\top}\bm x_S\right)^t}{t!}\,p_r\!\left(x_k-t;\bm\theta_k\right),\quad x_k\in\mathbb N_0.
\]

For the sample \(\mathcal D_N\), define the corresponding local log-likelihood by
\[
\ell_{k,r}\!\left(\bm\alpha_{k,S},\bm\theta_k\right)=\sum_{i=1}^N\log p_{k,r}\!\left(x_k^{(i)}\mid\bm x_S^{(i)};\bm\alpha_{k,S},\bm\theta_k\right).
\]

Write \(\bm\alpha=(\bm\alpha_{k,\operatorname{pa}_G(k)})_{k\in V}\), \(\bm\theta=(\bm\theta_k)_{k\in V}\), and \(\bm\eta_m=(\bm\alpha,\bm\theta)\). By the DAG factorization, the full log-likelihood is \(\ell_m(\bm\eta_m)=\sum_{k\in V}\ell_{k,\tau_k}\!\left(\bm\alpha_{k,\operatorname{pa}_G(k)},\bm\theta_k\right)\). 

Despite the node-wise decomposition, direct likelihood optimization remains costly: each local objective involves convolution and must be maximized repeatedly over candidate parent sets and exogenous families. The resulting computational burden grows rapidly with the number of nodes, motivating a plug-in BIC approach based on MoM estimates.

\subsection{MoM-based parameter estimation}
\label{MoMest}
We now construct the MoM-based estimates used to compute the local scores. Fix a node \(k\in V\) and a candidate parent set \(S\subseteq V\setminus\{k\}\). Let \(\bm\mu=\mathbb{E}[\bm X]\) and \(\bm\Sigma=\operatorname{Cov}(\bm X)\), with empirical counterparts \(\hat{\bm\mu}\) and \(\hat{\bm\Sigma}\), and use subscripts of \(\bm\mu\) and \(\bm\Sigma\) to denote components, subvectors, and submatrices. For a PT-SEM parameterization with parent set \(S\) at node \(k\), the conditional representation in Section~\ref{likelihood} gives
\[
\mathbb{E}[X_k\mid\bm X_S]=\mathbb{E}[\varepsilon_k]+\bm\alpha_{k,S}^{\top}\bm X_S,\quad \operatorname{Var}(X_k\mid\bm X_S)=\operatorname{Var}(\varepsilon_k)+\bm\alpha_{k,S}^{\top}\bm X_S.
\]
Taking covariance in the first identity yields \(\bm\Sigma_{Sk}=\bm\Sigma_{SS}\bm\alpha_{k,S}\).
By Lemma~\ref{lemma:cov_positive_definite}, \(\bm\Sigma_{SS}\) is positive definite for \(S \neq \varnothing\), so \(\hat{\bm\Sigma}_{SS}\) is nonsingular with probability tending to one. 

For any candidate \(S\neq\varnothing\), define the unconstrained MoM
\[
\tilde{\bm\alpha}_{k,S}=\{\tilde{\alpha}_{kj}\}_{j \in S} = 
\hat{\bm\Sigma}_{SS}^{-1}\hat{\bm\Sigma}_{Sk}
\]
and its nonnegative truncation
\[
\hat{\bm{\alpha}}_{k,S}= 
\left\{\hat{\alpha}_{kj}\right\}_{j \in S}, \quad 
\hat{\alpha}_{kj} = \max(\tilde\alpha_{kj},0), \; j \in S
\]
respectively. 
For \(S=\varnothing\), both \(\tilde{\bm\alpha}_{k,S}\) and \(\hat{\bm\alpha}_{k,S}\) are empty vectors.

The laws of total expectation and total variance express the exogenous mean and variance in terms of \(\bm\mu\), \(\bm\Sigma\), and \(\bm\alpha_{k,S}\). Replacing the population moments by their empirical counterparts and
\(\bm\alpha_{k,S}\) by \(\hat{\bm\alpha}_{k,S}\), yields 
\begin{equation}
\label{eq:exogenous-moment-estimates}
\hat m_k(S)=\hat\mu_k-\hat{\bm\alpha}_{k,S}^{\top}\hat{\bm\mu}_S, \quad 
\hat v_k(S)=\hat\Sigma_{kk}-\hat{\bm\alpha}_{k,S}^{\top}
\left(\hat{\bm\mu}_S+\hat{\bm\Sigma}_{SS}\hat{\bm\alpha}_{k,S}
\right).
\end{equation}

For each candidate exogenous family \(\mathcal F_r\), let \(H_r:\mathbb R^2\to\Theta_r\) be a fixed mapping that agrees with the moment inversion whenever that inversion is well-defined, and define
\[
\hat{\bm\theta}_{k,r}(S)=H_r\!\left(\hat m_k(S),\hat v_k(S)\right).
\]

For each $(k,S)$, $\hat{\bm\alpha}_{k,S}$, $\hat m_k(S)$, and $\hat v_k(S)$ do not depend on the candidate exogenous family $F_r$, and hence need to be computed only once.

For a candidate model \(m=(G, \bm\tau)\), collect the resulting MoM-based estimates as
\[
\hat{\bm\eta}^{\,\mathrm{MoM}}_m=\left((\hat{\bm\alpha}_{k,\operatorname{pa}_G(k)})_{k\in V},(\hat{\bm\theta}_{k,\tau_k}(\operatorname{pa}_G(k)))_{k\in V}\right).
\]

The following result establishes that the observed variables have finite moments of all orders, as needed for the asymptotic analysis below. 
\begin{lemma}
\label{lem:moment_finiteness_ptsem}
Under \hyperref[ass:local-mgf]{\textup{(A3)}}, for every node \(k\in V\), there exists \(t_k^\star>0\) such that \(\mathbb{E}[e^{t_k^\star X_k}]<\infty\). Consequently, \(X_k\) has finite moments of every order.
\end{lemma}

Let \(Z(\bm X)=\bigl(\bm X^\top,\operatorname{vech}(\bm X\bm X^\top)^\top\bigr)^\top\), \(\bm\nu=\mathbb{E}[Z(\bm X)]\), and \(\hat{\bm\nu}_N=N^{-1}\sum_{i=1}^N Z(\bm X^{(i)})\). Lemma~\ref{lem:moment_finiteness_ptsem} implies that \(\mathbb{E}\left[\Vert Z(\bm X)\Vert^2\right]<\infty\), so the multivariate central limit theorem gives
\[
\sqrt N\bigl(\hat{\bm\nu}_N-\bm\nu\bigr)\xrightarrow{d}\mathcal N(\bm 0,\bm\Omega),\quad \bm\Omega=\operatorname{Var}\{Z(\bm X)\}.
\]

For a correctly specified candidate model $m=(G,\tau)$, let
$\bm\eta_{m,0}\in\Theta_m$ denote a population parameter satisfying $P_{m,\bm\eta_{m,0}}=P_0$. 
In particular, a correctly specified candidate model may have a DAG that is a strict supergraph of $G_0$; in this case, the coefficients of the additional edges are zero at $\bm\eta_{m,0}$.

The parameter vector of a candidate exogenous family may contain discrete components; for example, the number of trials $n$ in the binomial family is integer-valued. Accordingly, we impose the following local regularity condition on the moment-inversion maps $H_r$. 

\begin{description}
\item[\textup{(A4)}]\phantomsection\label{ass:mom-regularity}
For each candidate family representing a true exogenous distribution, the continuous parameter components recovered by $H_r$ are $C^1$ functions of the mean and variance in a neighborhood of the corresponding population moment pair, while any discrete parameter component is locally constant there.
\end{description}

By Lemma~\ref{lemma:cov_positive_definite}, the relevant population covariance matrices are nonsingular. Together with the joint central limit theorem and \hyperref[ass:mom-regularity]{\textup{(A4)}}, this yields the following \(\sqrt{N}\)-consistency.

\begin{proposition}
\label{prop:plugin-rate}
For any correctly specified candidate model $m$, under \textup{(A4)},
\[
\bigl\|\hat{\bm\eta}^{\,\mathrm{MoM}}_m-\bm\eta_{m,0}\bigr\|=O_p(N^{-1/2}).
\]
\end{proposition}

\subsection{Model evaluation via BIC}
\label{subsec:bic}

We now define a plug-in BIC based on the MoM-based parameter estimates for candidate-model evaluation and structure learning in PT-SEM, and establish consistency of the resulting plug-in BIC for DAG and exogenous-family selection. It evaluates each candidate likelihood at the MoM-based estimates from Section~\ref{MoMest}, avoiding repeated numerical maximization while retaining the usual BIC dimension penalty.

For a candidate model \(m=(G,\bm\tau)\), let \(\Theta_m\) denote its parameter space. For \(r\in\mathcal I\), let \(q_r\) denote the number of continuous free parameters in \(\mathcal F_r\), and define \(q(m)=|E|+\sum_{k\in V}q_{\tau_k}\). The MoM plug-in BIC is defined as
\[
\operatorname{BIC}^{\mathrm{MoM}}(m)=-2\ell_m\!\left(\hat{\bm\eta}^{\,\mathrm{MoM}}_m\right)+q(m)\log N.
\]

For \(k\in V\), \(S\subseteq V\setminus\{k\}\), and \(r\in\mathcal I\), define
\[
s_{k,r}(S)=-2\ell_{k,r}\!\left(\hat{\bm\alpha}_{k,S},\hat{\bm\theta}_{k,r}(S)\right)+\bigl(|S|+q_r\bigr)\log N,\quad s_k(S)=\min_{r\in\mathcal I}s_{k,r}(S),
\]
\[
\operatorname{BIC}^{\mathrm{MoM}}(G)=\min_{\bm\tau\in\mathcal I^V}\operatorname{BIC}^{\mathrm{MoM}}\!\bigl((G,\bm\tau)\bigr)=\sum_{k\in V}s_k\!\left(\operatorname{pa}_G(k)\right).
\]
Thus, \(\operatorname{BIC}^{\mathrm{MoM}}(G)\) retains the node-wise decomposability required for structure search.

For the consistency analysis, in addition to \hyperref[ass:mom-regularity]{\textup{(A4)}}, we impose the following regularity condition on the candidate exogenous families.
For the compactness requirement below, we restrict each parameter space to a sufficiently large compact subset containing the relevant population parameters; this restriction is introduced only for the theoretical analysis.
\begin{description}
\item[\textup{(A5)}]\phantomsection\label{ass:bic-regularity} 
For each candidate family $\mathcal F_r$, the parameter space $\Theta_r$ is compact, with each discrete parameter component taking values in a finite set, and $p_r(x;\bm\theta)$ is continuous in its continuous parameters. At every correctly specified parameter $\bm\theta_0$, the continuous parameter components are interior, and the support is locally fixed. Moreover, there exists a neighborhood $\mathcal N_r$ of $\bm\theta_0$ such that, uniformly over $\bm\theta\in\mathcal N_r$, the first two derivatives of $\log p_r(x;\bm\theta)$ with respect to the continuous parameters and the ratios \(p_r(x-h;\bm\theta)/p_r(x;\bm\theta)\), \(h=1,2\), are bounded in absolute value by a polynomial in $x$ on the common
support. 
\end{description}

Condition \hyperref[ass:bic-regularity]{\textup{(A5)}} provides the global and local regularity used in the consistency proof. Compactness and continuity separate misspecified candidates from the true distribution, while the derivative and mass-ratio bounds control tail growth and yield polynomial envelopes for the score and Hessian of the Poisson--exogenous convolution. Lemma~\ref{lem:moment_finiteness_ptsem} then ensures the required integrability; see Appendix~\ref{app:proofs-section4}. These conditions are stated in a sufficient form rather than a minimal one, so that the required likelihood bounds can be checked directly for concrete candidate families.

Let \(\mathcal T_0\subseteq\mathcal I^V\) be the set of family assignments that represent the true node-wise exogenous distributions and minimize \(\sum_{k\in V}q_{\tau_k}\) among all such assignments. Fix any \(\bm\tau_0\in\mathcal T_0\), and write \(m_0=(G_0,\bm\tau_0)\) and \(\bm\eta_0=\bm\eta_{m_0,0}\).

\begin{proposition}
\label{prop:mom-bic-gap}
Let $\ell_0=\ell_{m_0}(\bm\eta_0)$. Under
\hyperref[ass:mom-regularity]{\textup{(A4)}} and
\hyperref[ass:bic-regularity]{\textup{(A5)}}, \(\ell_{m_0}\!\left(\hat{\bm\eta}^{\,\mathrm{MoM}}_{m_0}\right)=\ell_0+O_p(1)\). For every correctly specified candidate model $m$, \(\ell_m\!\left(\hat{\bm\eta}^{\,\mathrm{MoM}}_m\right)\le \ell_0+O_p(1).\)
\end{proposition}

For any correctly specified candidate model \(m=(G,\bm\tau)\) with \(G\neq G_0\) or \(\bm\tau\notin\mathcal T_0\), PT-SEM identifiability and the definition of \(\mathcal T_0\) imply \(q(m)\ge q(m_0)+1\): such a model either contains a redundant edge or uses a nonminimal correct family assignment. Proposition~\ref{prop:mom-bic-gap} shows that its plug-in likelihood can exceed the true likelihood by at most \(O_p(1)\). Hence the additional \(\log N\) penalty eliminates every such correctly specified candidate.

For a candidate model that cannot represent the true distribution, Appendix~\ref{app:proofs-section4} establishes a strictly positive population likelihood gap, yielding an \(O(N)\) likelihood loss that dominates the \(O(\log N)\) difference in dimension penalties. Since the candidate model collection is finite, these two cases yield joint consistency of DAG and family selection.

\begin{theorem}
\label{thm:bic-consistency}
Let \((\hat G,\hat{\bm\tau})\in\arg\min_{G\in\mathcal G,\,\bm\tau\in\mathcal I^V}\operatorname{BIC}^{\mathrm{MoM}}\!\bigl((G,\bm\tau)\bigr)\). Under \hyperref[ass:zero-mass]{\textup{(A1)}}--\hyperref[ass:bic-regularity]{\textup{(A5)}},
\[
\Pr\!\left(\hat G=G_0,\hat{\bm\tau}\in\mathcal T_0\right)\to1 \quad \textit{as}\quad  N\to\infty.
\]
\end{theorem}
\subsection{Structure learning by exact dynamic programming}
\label{subsec:dp}

The profiled local scores \(\{s_k(S)\}_{k,S}\) preserve the decomposable form of \(\operatorname{BIC}^{\mathrm{MoM}}(G)\), allowing us to apply the exact subset dynamic programming (DP) algorithm of \citet{SilanderMyllymaki2006UAI}. The resulting search proceeds over node subsets rather than by explicit enumeration of DAGs and yields a global minimizer of \(\operatorname{BIC}^{\mathrm{MoM}}(G)\) over \(G\in\mathcal G\).

Algorithm~\ref{alg:ptsem-learning} summarizes the resulting two-stage procedure. Stage 1 precomputes the local scores and stores the coefficient estimates and a minimizing family label for each node--parent-set pair. Stage 2 applies the exact subset-DP and backtracking procedure of \citet{SilanderMyllymaki2006UAI}, denoted by \(\operatorname{DP}\) below, to obtain \(\hat G\); the stored quantities then recover the thinning coefficients and family assignment. The overall time complexity is \(O\!\left((d^4+Nd^2)2^d\right)\), with space complexity \(O(d^2 2^d)\).

\begin{algorithm}[ht]
\caption{PT-SEM learning by DP-BIC}
\label{alg:ptsem-learning}
\DontPrintSemicolon
\SetInd{0.5em}{1.0em}
\KwIn{sample \(\mathcal D_N\), node set \(V\), and candidate exogenous families \(\mathcal M=\{\mathcal F_r:r\in\mathcal I\}\)}
\KwOut{DAG \(\hat G\), thinning coefficients \(\hat{\bm\alpha}_{\hat G}\), and family assignment \(\hat{\bm\tau}\)}
\tcc{Stage 1: Local-score precomputation}
\ForEach{\(k\in V\), \(S\subseteq V\setminus\{k\}\)}{
compute \(\hat{\bm\alpha}_{k,S}\), \(\hat m_k(S)\), and \(\hat v_k(S)\);\;
\ForEach{\(r\in\mathcal I\)}{
set \(\hat{\bm\theta}_{k,r}(S)\leftarrow H_r\!\left(\hat m_k(S),\hat v_k(S)\right)\);\;
evaluate \(\ell_{k,r}\!\left(\hat{\bm\alpha}_{k,S},\hat{\bm\theta}_{k,r}(S)\right)\) and compute \(s_{k,r}(S)\);\;
}
choose \(\hat r_k(S)\in\arg\min_{r\in\mathcal I}s_{k,r}(S)\) and set \(s_k(S)\leftarrow s_{k,\hat r_k(S)}(S)\);\;
}
\tcc{Stage 2: Exact graph optimization and parameter recovery}
\(\hat G\leftarrow\operatorname{DP}\!\left(\{s_k(S)\}_{k,S}\right)\);\;
\(\hat{\bm\alpha}_{\hat G}\leftarrow\left(\hat{\bm\alpha}_{k,\operatorname{pa}_{\hat G}(k)}\right)_{k\in V}\);\;
\(\hat{\bm\tau}\leftarrow\left(\hat r_k\!\left(\operatorname{pa}_{\hat G}(k)\right)\right)_{k\in V}\);\;
\end{algorithm}

\section{Numerical Experiments}
\label{sec:experiments}
We evaluate the proposed structure-learning procedure, which combines local plug-in BIC scoring with exact dynamic programming, on synthetic data generated from PT-SEMs. We assess DAG recovery, thinning-coefficient estimation, and the finite-sample behavior of node-wise exogenous-family selection.

\subsection{Simulation design}
\label{subsec:simulation-design}
We randomly generate DAGs with a fixed number of nodes $d$ and a fixed average in-degree $\bar{k}_{\mathrm{in}}$.
We consider two exogenous-distribution settings. In the all-Poisson setting, all exogenous distributions are Poisson. In the \(\mathcal M\) setting, the exogenous family at each node is drawn independently and uniformly from
\[
\mathcal M=\{\mathrm{Poisson},\mathrm{NB},\mathrm{ZIP},\mathrm{Geometric},\mathrm{Binomial},\mathrm{Bernoulli}\},
\]
where NB and ZIP denote the negative binomial and zero-inflated Poisson distributions, respectively. 
In the all-Poisson setting, the Poisson parameter at each node is drawn from $\operatorname{Uniform}(2,10)$. 
In the \(\mathcal M\) setting, the distribution parameters are drawn according to the corresponding row of Table~\ref{tab:exogenous-parameters}.

\begin{table}[ht]
\centering
\caption{Parameter settings for the exogenous distributions.}
\label{tab:exogenous-parameters}
\small
\setlength{\tabcolsep}{4pt}
\renewcommand{\arraystretch}{1.08}
\begin{tabular}{@{}p{0.23\linewidth}p{0.71\linewidth}@{}}
\hline
Family & Parameter setting \\
\hline
Poisson &
\(\lambda_i\sim\operatorname{Uniform}(2,10)\) \\
Negative binomial &
\(r_i\sim\operatorname{Uniform}(2,10)\), \(p_i\sim\operatorname{Uniform}(0.15,0.85)\) \\
Zero-inflated Poisson &
\(\lambda_i\sim\operatorname{Uniform}(2,10)\), \(\rho_i\sim\operatorname{Uniform}(0.15,0.85)\) \\
Geometric &
\(\mu_i\sim\operatorname{Uniform}(2,10)\), \(p_i=(1+\mu_i)^{-1}\) \\
Binomial &
\(n_i\) is drawn uniformly from \(\{2,\ldots,20\}\), \(p_i\sim\operatorname{Uniform}(0.15,0.85)\) \\
Bernoulli &
\(p_i\sim\operatorname{Uniform}(0.15,0.85)\) \\
\hline
\end{tabular}
\end{table}

We refer to the proposed method as Proposed DP-BIC and compare it with benchmarks:
\begin{itemize}
\setlength{\parskip}{0cm} 
  \setlength{\itemsep}{0cm} 
    \item Greedy-BIC: A variant of the proposed method that starts from the empty DAG and repeatedly applies the legal single-edge addition, deletion, or reversal that yields the largest decrease in the plug-in BIC score, until no such move improves the score.
    \item Poisson-only DP: The proposed method under the restriction that all exogenous distributions are Poisson.
     \item Oracle DP: The proposed method with the exogenous distribution family at each node known, while its parameters remain unknown.
    \item ODS: The structure-learning method for Poisson DAGs proposed by \citet{park2015learning}. The all-Poisson PT-SEM considered below is an identity-link Poisson DAG.
    \item PC-RCIT: The PC algorithm \citep{spirtes2001causation} combined with the randomized conditional independence test \citep{strobl2019}.
    \item PB-SCM: The higher-order-cumulant-based structure-learning method for PB-SCM proposed by \citet{qiao2024causal}.
    \item PB-SCM-PGF: The probability-generating-function-based structure-learning method for PB-SCM proposed by \citet{NEURIPS2024_15aaa922}.
\end{itemize}

We consider two thinning-coefficient regimes. In the restricted regime, the coefficients are independently sampled from $\operatorname{Uniform}(0.15,0.85)$, and PB-SCM and PB-SCM-PGF are included as benchmarks. In the extended regime, the coefficients are independently sampled from $\operatorname{Uniform}(0.2,2.0)$, and these two methods are not included. Because all data sets are generated from PT-SEM, the simulation design naturally favors the proposed method. 
Note, however, that the all-Poisson setting also lies within the Poisson DAG model class.

For each exogenous-distribution setting and thinning-coefficient regime, we conduct three sweeps over the dimension, sample size, and average in-degree, with the corresponding settings summarized in Table~\ref{tab:sim-design}.
\begin{table}[ht]
\centering
\caption{Regime and sweep settings}
\label{tab:sim-design}
\small
\setlength{\tabcolsep}{4pt}
\renewcommand{\arraystretch}{1.08}
\begin{tabular}{@{}p{0.28\linewidth}p{0.66\linewidth}@{}}
\hline
Item  & Setting \\
\hline
Dimension sweep &
\(d\in \{4,5,6,7,8,9,10\}\), \(N=3200\), \(\bar k_{\mathrm{in}}=1.5\) \\
Sample size sweep &
\(N\in\{100,200,400,800,1600,3200,6400,10000\}\), \(d=8\), \(\bar k_{\mathrm{in}}=1.5\) \\
Average in-degree sweep &
\(\bar k_{\mathrm{in}}\in\{1.0,1.5,2.0,2.5,3.0\}\), \(d=8\), \(N=3200\) \\
\hline
\end{tabular}
\end{table}

For the six exogenous families used in the simulations, Table~\ref{tab:simulation-family-regularity} summarizes the moment inversions relevant to \hyperref[ass:mom-regularity]{\textup{(A4)}} and the mass-ratio orders relevant to \hyperref[ass:bic-regularity]{\textup{(A5)}}, where \(m=\mathbb E[\varepsilon]\) and \(v=\operatorname{Var}(\varepsilon)\). The last column gives the order of \(p_r(x-h;\bm\theta)/p_r(x;\bm\theta)\) for \(h=1,2\). The inversion formulas are understood on the regions where they are well defined.

\begin{table}[ht]
\centering
\caption{Moment inversions and mass-ratio bounds for the candidate exogenous families.}
\label{tab:simulation-family-regularity}
\small
\setlength{\tabcolsep}{4pt}
\renewcommand{\arraystretch}{1.08}
\begin{tabular}{@{}p{0.22\linewidth}p{0.59\linewidth}p{0.13\linewidth}@{}}
\hline
Family & Moment inversion \(H_r(m,v)\) & Mass-ratio order \\
\hline
Poisson & \(\lambda=m\) & \(O((1+x)^h)\) \\
Negative binomial & \(r=m^2/(v-m),\ p=m/v\) & \(O(1)\) \\
Zero-inflated Poisson & \(\rho=(v-m)/(v-m+m^2),\ \lambda=(v-m+m^2)/m\) & \(O((1+x)^h)\) \\
Geometric & \(p=(1+m)^{-1}\) & \(O(1)\) \\
Binomial & \(n=\operatorname{round}\{m^2/(m-v)\},\ p=m/n\) & \(O(1)\) \\
Bernoulli & \(p=m\) & \(O(1)\) \\
\hline
\end{tabular}
\end{table}

For \(x<h\), \(p_r(x-h;\bm\theta)\) is interpreted as zero. The inversion maps in Table~\ref{tab:simulation-family-regularity} are continuously differentiable in their continuous coordinates near the generating moment pairs, while the rounded Binomial trial number is locally constant there; hence \hyperref[ass:mom-regularity]{\textup{(A4)}} holds. For \hyperref[ass:bic-regularity]{\textup{(A5)}}, take the parameter spaces to be compact with the generating values in the interiors of their continuous coordinates. The supports are then locally fixed, and the displayed mass-ratio bounds and the first two derivatives of \(\log p_r(x;\bm\theta)\) admit uniform polynomial bounds in \(x\).

We evaluate DAG estimation accuracy using the directed-edge F1 score. Let $E_0$ and $\hat{E}$ denote the directed-edge sets of the true and estimated graphs, respectively. Define the set of correctly recovered edges by $\mathcal{C}=E_0\cap\hat{E}$. When $\mathcal{C}\neq\varnothing$, we evaluate coefficient estimation accuracy by the mean absolute percentage error (MAPE) over the correctly recovered edges:
\[
100\cdot
\frac{1}{|\mathcal C|}
\sum_{(j,i)\in \mathcal C}
\frac{|\hat{\alpha}_{ij}-\alpha_{ij}|}{\alpha_{ij}}.
\]
Each simulation configuration uses $R=100$ independent replications. Curves show means across replications, and shaded bands indicate $\pm 1.96$ standard errors.
\subsection{Results}
\label{subsec:directed-recovery}

\begin{figure}[ht]
\centering
\includegraphics[width=1\textwidth]{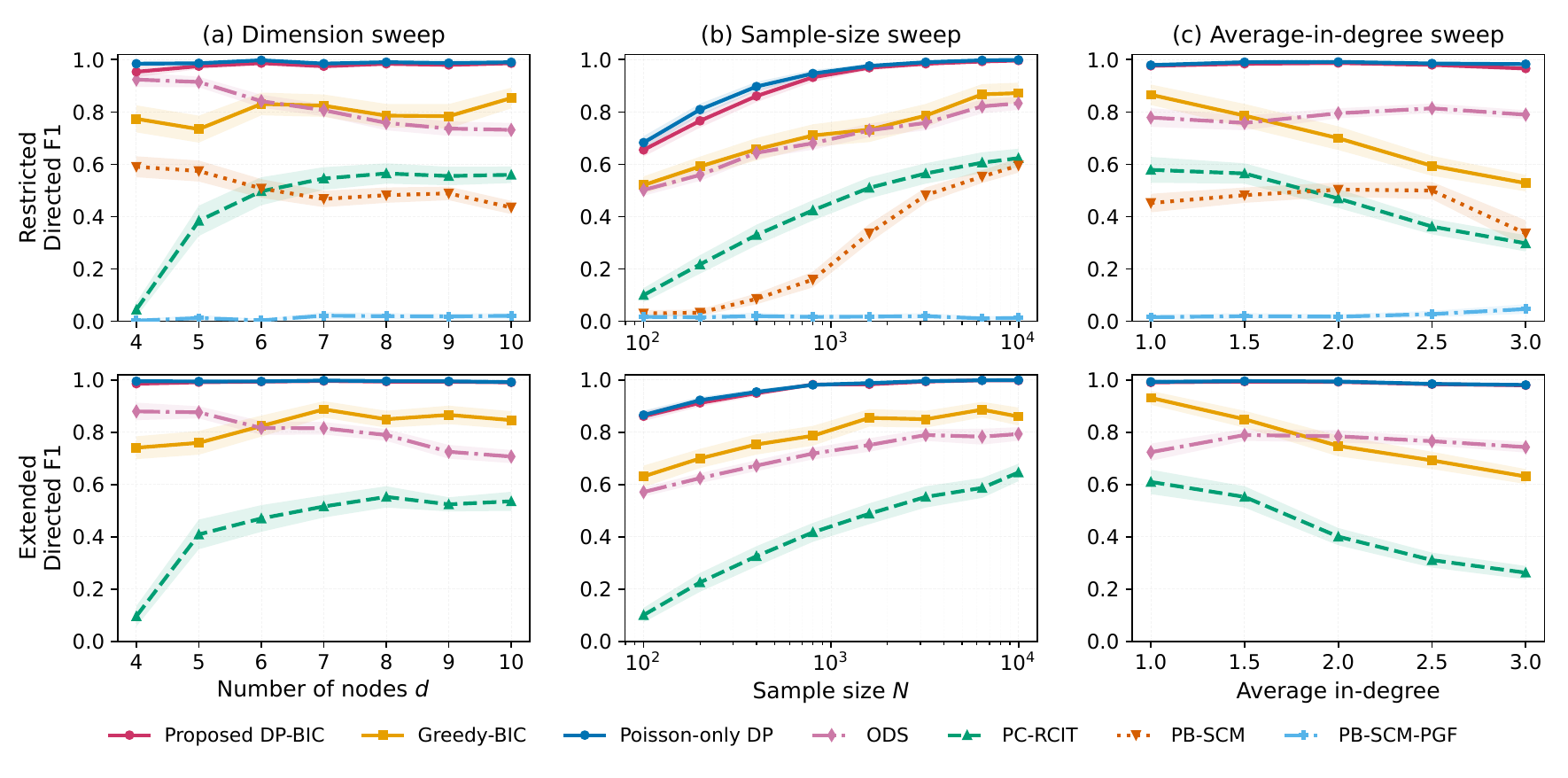}
\caption{Directed-edge F1 in the all-Poisson setting.}
\label{fig:all-poisson-overlap}
\end{figure}

We first consider the all-Poisson setting, which is also an identity-link Poisson DAG and therefore provides a natural baseline for comparison. Figure~\ref{fig:all-poisson-overlap} reports the F1 scores for the dimension, sample-size, and average-in-degree sweeps under both the restricted and extended regimes. Poisson-only DP and Proposed DP-BIC achieve similar DAG-recovery accuracy throughout, with their difference decreasing as the sample size increases. 
Proposed DP-BIC also consistently outperforms ODS across the settings considered. These results indicate that profiling over the six candidate exogenous families incurs little loss in DAG recovery.

\begin{figure}[!hb]
\centering
\includegraphics[width=1\textwidth]{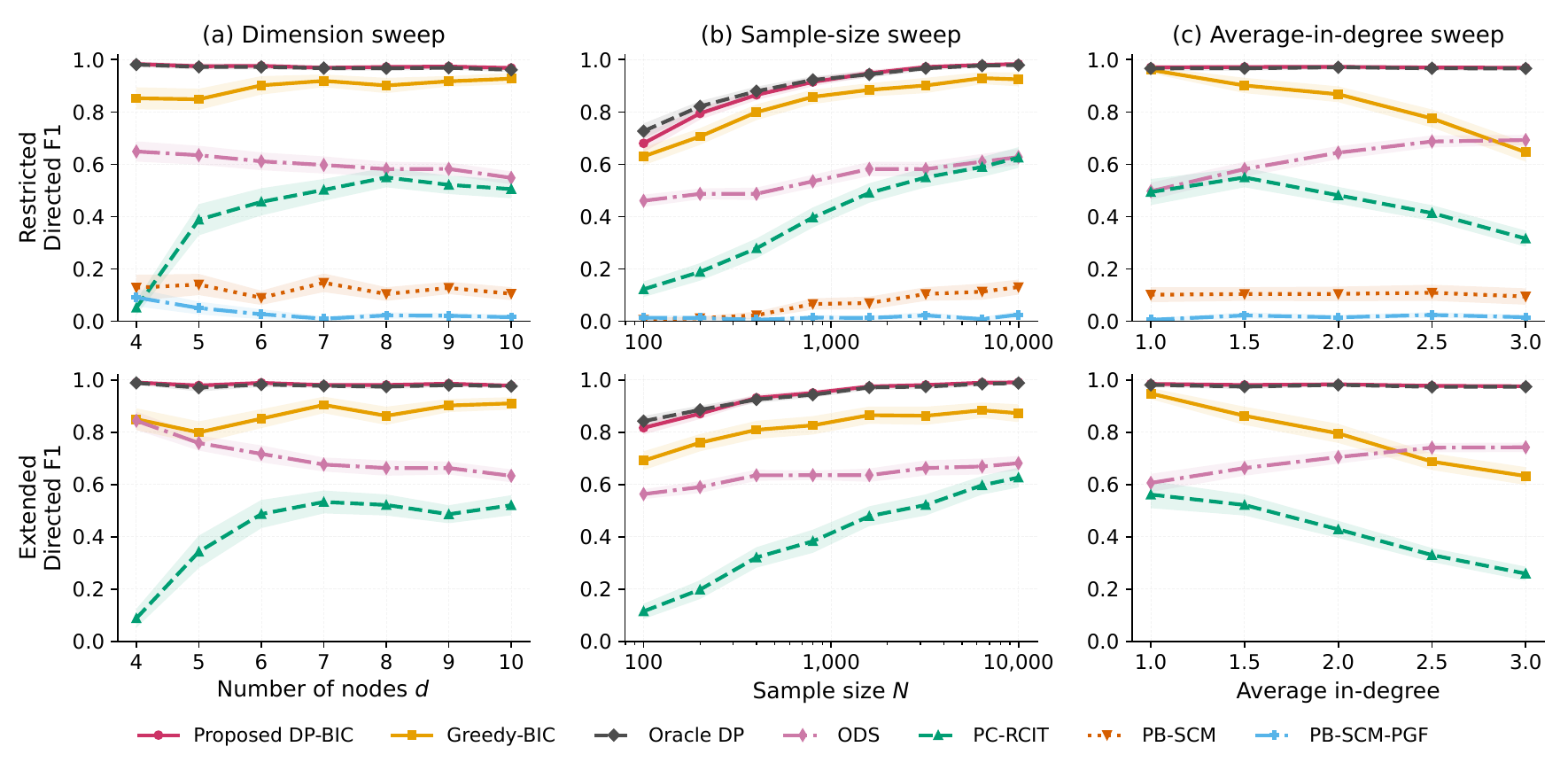}
\caption{Directed-edge F1 in the $\mathcal M$ setting.}
\label{fig:directed-f1}
\end{figure}

We next consider the $\mathcal M$ setting, in which the exogenous distribution family may vary across nodes. Figure~\ref{fig:directed-f1} reports the F1 scores for the three simulation sweeps under both the restricted and extended regimes. Proposed DP-BIC performs nearly as well as Oracle DP under both regimes, indicating little loss from data-driven exogenous-family selection. Its F1 score increases with sample size and remains high across the considered dimensions and average in-degrees. Proposed DP-BIC also outperforms Greedy-BIC and ODS; notably, the gap from Greedy-BIC widens as the average in-degree increases, showing the benefit of exact DP optimization. PC-RCIT yields lower directed F1 scores, consistent with recovery only up to a Markov equivalence class. Under the restricted regime, PB-SCM and PB-SCM-PGF achieve substantially lower recovery accuracy.

\begin{figure}[!hb]
\centering
\includegraphics[width=1\textwidth]{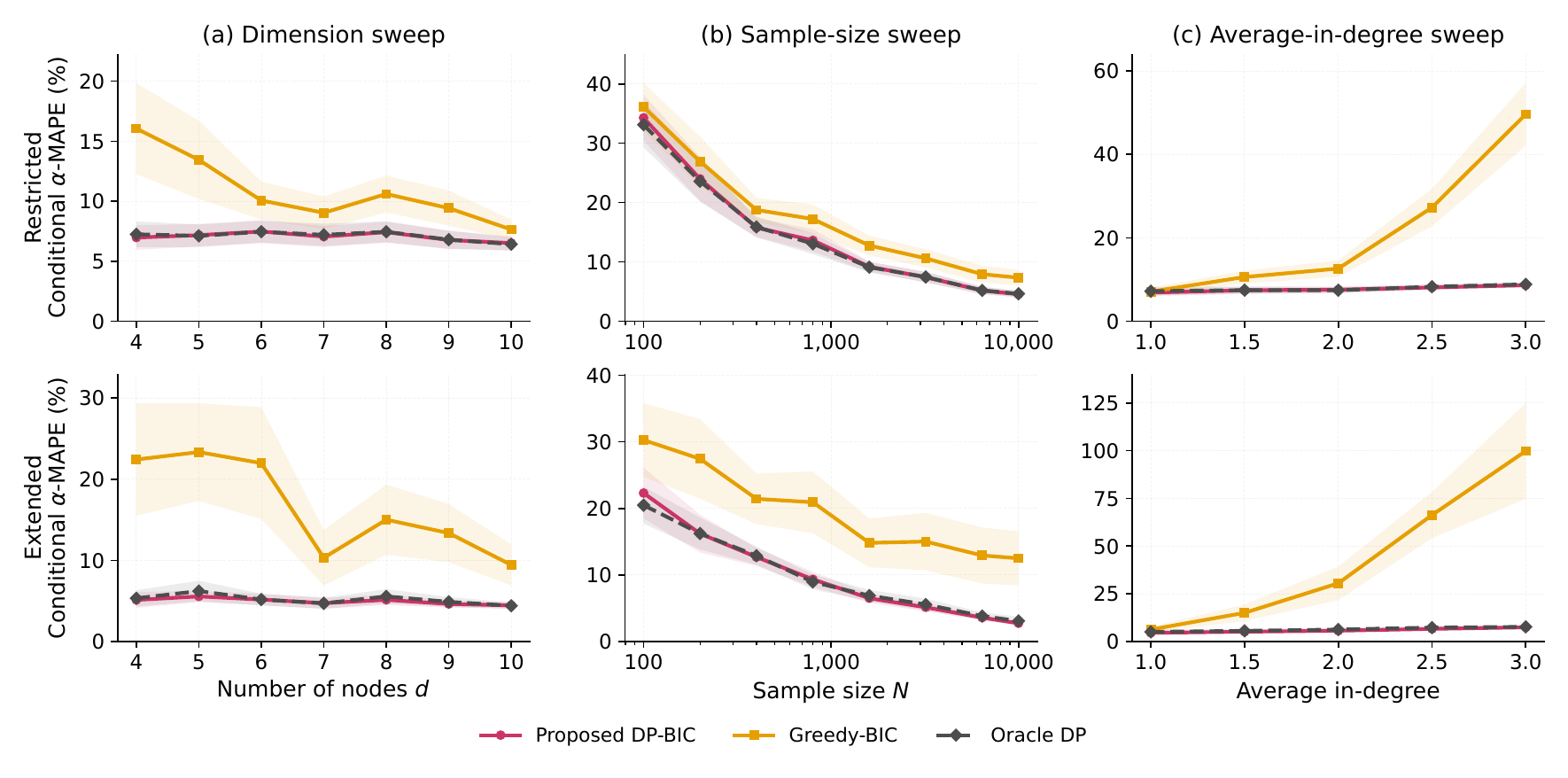}
\caption{Thinning-coefficient MAPE over correctly recovered edges in the $\mathcal M$ setting.}
\label{fig:conditional-alpha-error}
\end{figure}

Figure~\ref{fig:conditional-alpha-error} reports the MAPE of the estimated thinning coefficients over correctly recovered edges for Proposed DP-BIC, Oracle DP, and Greedy-BIC. Under both regimes, Proposed DP-BIC closely tracks Oracle DP, and the MAPE decreases as sample size increases. As the average in-degree increases, Greedy-BIC deteriorates markedly, whereas Proposed DP-BIC remains stable and close to Oracle DP.

\begin{figure}[!t]
\centering
\includegraphics[width=1\textwidth]{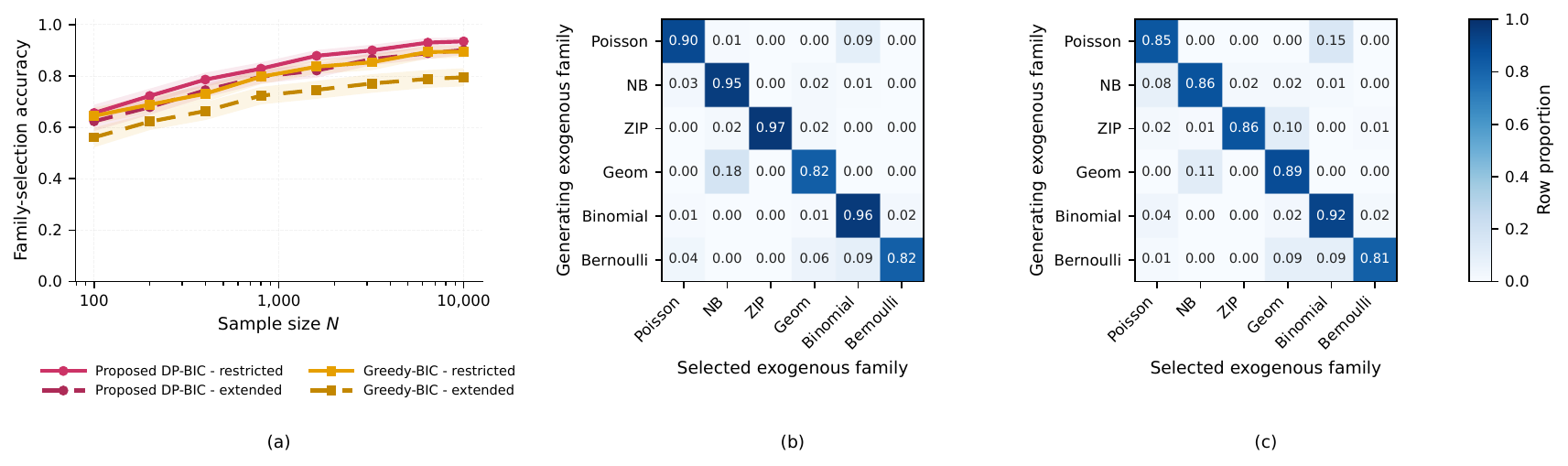}
\caption{Exogenous-family selection diagnostics in the $\mathcal M$ setting. Panel~(a) shows family-selection accuracy along the sample-size sweep. Panels~(b) and~(c) show row-normalized confusion matrices for Proposed DP-BIC under the restricted and extended regimes, respectively.}
\label{fig:working-family-diagnostics}
\end{figure}

Figure~\ref{fig:working-family-diagnostics} summarizes the proportion of nodes for which the generating exogenous family is correctly selected from the six candidate families in $\mathcal M$. Figure~\ref{fig:working-family-diagnostics}(a) plots this proportion against the sample size for Proposed DP-BIC and Greedy-BIC under the restricted and extended regimes. Figures~\ref{fig:working-family-diagnostics}(b) and (c) present the corresponding confusion matrices under the restricted and extended regimes, for the setting with $d=8$, $N=3200$, and $\bar{k}_{\mathrm{in}}=1.5$.

The results show that the proportion of correct selections is generally high when the sample size is sufficiently large and is higher under the restricted regime than under the extended regime. Although exogenous-family selection is less accurate when $N$ is around 100, Figure~\ref{fig:directed-f1} shows that the corresponding DAG recovery accuracy remains nearly identical to that of Oracle DP. This suggests that DAG recovery is fairly robust to errors in exogenous-family selection.
\section{Real-World Event-Count Experiment}
\label{sec:real-world}
We evaluate PT-SEM on NBA play-by-play event counts derived from version 8 of the Kaggle data set \emph{NBA WNBA play-by-play and shots data}, compiled by \citet{shufinskiy2025nba}. 
We use data from the 2015--16 through 2024--25 NBA regular seasons. Events are aggregated at the team-quarter level, with overtime excluded, yielding 95,808 observations.

We consider five event counts for the focal team in each quarter. FOUL counts non-offensive fouls drawn by the team, FTA and FTM count its free-throw attempts and made free throws, and PERS and LOOSE count the subsets of drawn fouls recorded as personal and loose-ball fouls, respectively. Under the NBA definitions, personal fouls involve illegal physical contact with an opponent, 
whereas loose-ball fouls occur when neither team has control of the ball, such as in a rebound contest~\citep{nbaRulebook}.

\begin{figure}[ht]
\centering
\includegraphics[width=0.45\textwidth]{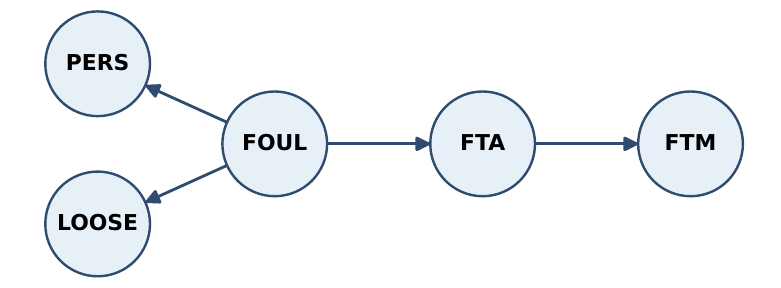}
\caption{Prespecified reference DAG for the NBA event counts.}
\label{fig:nba-five-node-benchmark}
\end{figure}

We construct the reference DAG in Figure~\ref{fig:nba-five-node-benchmark} from the event-count definitions above and the official NBA rulebook \citep{nbaRulebook}. FOUL counts the non-offensive foul events drawn by the focal team, while PERS and LOOSE record two corresponding foul subtypes and FTA records free-throw attempts awarded from qualifying foul events. FTM records successful free-throw attempts. Accordingly, the prespecified reference DAG takes FOUL as the common parent of PERS, LOOSE, and FTA, and FTA as the parent of FTM.

We fit PT-SEM using Proposed DP-BIC with the same six candidate exogenous families as in the simulations and compare it with ODS, PC-RCIT, PB-SCM, and PB-SCM-PGF. Each method is fitted separately to each season and evaluated against the reference DAG in Figure~\ref{fig:nba-five-node-benchmark}. Precision, recall, and F1 are computed separately for the skeleton and the directed edge set and then averaged across the ten seasons. Skeleton metrics ignore edge orientation, whereas directed-edge metrics require both adjacency and orientation to agree with the reference DAG. 

\begin{table}[ht]
\centering
\small
\caption{Average structural-recovery performance over the ten NBA seasons.}
\label{tab:nba-main}
\begin{tabular}{lcccccc}
\toprule
Method
& \multicolumn{3}{c}{Skeleton recovery}
& \multicolumn{3}{c}{Directed-edge recovery} \\
\cmidrule(lr){2-4}\cmidrule(lr){5-7}
& Prec. & Rec. & F1 & Prec. & Rec. & F1 \\
\midrule
Proposed DP-BIC
& 1.000 & 1.000 & 1.000
& 1.000 & 1.000 & 1.000 \\
ODS
& 0.800 & 1.000 & 0.889
& 0.800 & 1.000 & 0.889 \\
PC-RCIT
& 0.619 & 1.000 & 0.764
& 0.148 & 0.200 & 0.169 \\
PB-SCM
& 0.900 & 0.575 & 0.688
& 0.650 & 0.300 & 0.399 \\
PB-SCM-PGF
& 0.719 & 0.925 & 0.806
& 0.423 & 0.450 & 0.423 \\
\bottomrule
\end{tabular}
\end{table}

Table~\ref{tab:nba-main} shows that Proposed DP-BIC recovers the complete reference DAG in every season, yielding precision, recall, and F1 of \(1.000\) for both skeleton and directed-edge recovery. ODS recovers all reference edges but includes additional edges, resulting in precision and F1 of \(0.800\) and \(0.889\), respectively, for both skeleton and directed-edge recovery. PC-RCIT attains perfect skeleton recall but substantially lower directed-edge recovery. PB-SCM and PB-SCM-PGF also yield markedly lower directed-edge recovery than Proposed DP-BIC.

\begin{figure}[!ht]
\centering
\includegraphics[width=\textwidth]{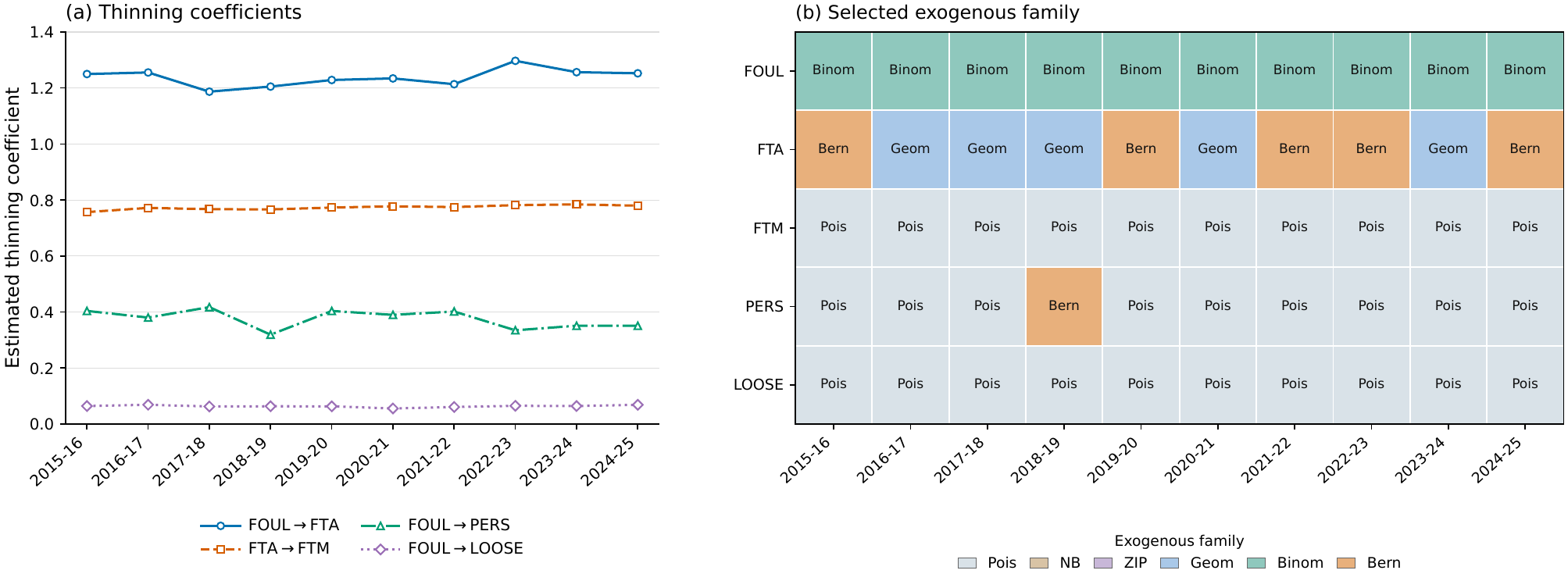}

\begin{minipage}[t]{0.48\textwidth}
\centering
\end{minipage}
\hfill
\begin{minipage}[t]{0.48\textwidth}
\centering
\end{minipage}

\caption{Season-wise PT-SEM estimates for the NBA data.}
\label{fig:nba-diagnostics}
\end{figure}

Figure~\ref{fig:nba-diagnostics} complements these aggregate results with season-wise PT-SEM estimates. Panel~(a) reports the estimated thinning coefficients for the reference edges, and Panel~(b) reports the exogenous family selected at each node in each season.
The estimated thinning coefficients are stable across seasons. In particular, the coefficient for FOUL\(\to\)FTA ranges from \(1.187\) to \(1.297\), with a mean of \(1.238\), and remains above one in every season. This magnitude is consistent with the fact that a qualifying foul can generate multiple free-throw attempts. The selected exogenous families vary across nodes and seasons, whereas the recovered DAG coincides with the reference DAG in every season.

Overall, Proposed DP-BIC recovers the prespecified NBA event relations more accurately than the competing methods, while producing stable and interpretable thinning-coefficient estimates and allowing node-wise heterogeneous exogenous-family assignments.
\section{Conclusion}
\label{sec:conclusion}

We introduced PT-SEM, a structural equation model for count-valued variables based on Poisson thinning. The thinning coefficients enter linearly in the conditional means and range over $[0,\infty)$, while the mutually independent exogenous distributions may vary across nodes. No parametric family specification for the exogenous distributions is required for identification. We showed that, under the stated node-wise conditions, the observational distribution uniquely determines the DAG, the thinning coefficient matrix, and the exogenous distributions. The proof uses a recursive sink characterization in terms of conditional cumulant generating functions. For binomial thinning, we showed that a nonsink can satisfy the sink characterization only when its exogenous distribution is Poisson; the all-Poisson PB-SCM lies in this exceptional case and is not fully identifiable in general.

For structure learning, we developed a decomposable plug-in BIC based on local moment estimates, avoiding repeated numerical maximization of local likelihoods. The score can be optimized exactly by subset dynamic programming, and the resulting procedure is consistent for DAG and exogenous-family selection. In simulations, the proposed DP-BIC compared favorably with the benchmark methods in both DAG recovery and thinning-coefficient estimation while selecting the exogenous family separately at each node. In a real-data analysis of multivariate event counts, the learned graphs more closely matched the reference relations derived from NBA event definitions and rules than those obtained by the competing methods, and the thinning-coefficient estimates were stable across seasons.

The current theory assumes causal sufficiency and acyclicity. Extensions to latent confounding and feedback systems, scalable score optimization for larger graphs, and models for mixed count, categorical, and continuous data remain open.



\acks{This work was supported by JST SPRING under Grant Number JPMJSP2110 and JSPS KAKENHI under Grant Number 25K15017.}




\appendix
\renewcommand{\thesection}{\Alph{section}}
\renewcommand{\thetheorem}{\thesection.\arabic{theorem}}
\section{Proofs of Lemmas and Theorems}
\label{sec:appendix}
\subsection{Proofs for Section~\ref{sec:model}}
\label{sec:proof sec 2}
\subsubsection{Proof of Proposition~\ref{prop:ptsem-markov}}
Let \(v_1,\ldots,v_d\) be a topological ordering of \(G\). For each \(m\), all parents of \(v_m\) belong to \(\{v_1,\ldots,v_{m-1}\}\). By the recursive PT-SEM construction, \(X_{v_m}\) depends on the preceding variables only through \(\bm X_{\operatorname{pa}_G(v_m)}\), while the exogenous noise \(\varepsilon_{v_m}\) and the Poisson offspring variables associated with the edges entering \(v_m\) are independent of the variables generated earlier. Therefore, 
\[
p_{X_{v_m}\mid X_{v_1},\ldots,X_{v_{m-1}}}
\left(x_{v_m}\mid x_{v_1},\ldots,x_{v_{m-1}}\right)
=
p_{X_{v_m}\mid\bm X_{\operatorname{pa}_G(v_m)}}
\left(x_{v_m}\mid \bm x_{\operatorname{pa}_G(v_m)}\right).
\]
Applying the chain rule in the topological order gives, for every \(\bm x\in\mathbb N_0^d\),
\[
\begin{aligned}
p_{\bm X}(\bm x)
&=
\prod_{m=1}^d
p_{X_{v_m}\mid X_{v_1},\ldots,X_{v_{m-1}}}
\left(x_{v_m}\mid x_{v_1},\ldots,x_{v_{m-1}}\right)\\
&=
\prod_{m=1}^d
p_{X_{v_m}\mid\bm X_{\operatorname{pa}_G(v_m)}}
\left(x_{v_m}\mid \bm x_{\operatorname{pa}_G(v_m)}\right)\\
&=
\prod_{i\in V}
p_{X_i\mid\bm X_{\operatorname{pa}_G(i)}}
\left(x_i\mid \bm x_{\operatorname{pa}_G(i)}\right).
\end{aligned}
\]
Thus the observational distribution factorizes according to \(G\) and is Markov.
\hfill\qed
\subsection{Proofs for Section~\ref{sec:identifiability}}
\label{app:proofs-section3}
\subsubsection{Proof of Lemma~\ref{lem:bivariate-three-layer-bayes-tilt}}
\label{proof:bivariate-three-layer-bayes-tilt}
Consider the bivariate PT-SEM in~\eqref{eq:bivariate-ptsem} and
write \(p_r:=\Pr(\varepsilon_2=r)\). Since \(X_1=\varepsilon_1\), the forward model gives
\[
\Pr(X_1=x,X_2=0)=\Pr(\varepsilon_1=x)p_0e^{-\alpha x}.
\]
Because \(p_0>0\), the conditional distribution \(X_1 \mid X_2=0\) has the same support as \(\varepsilon_1\) and is therefore nondegenerate by \hyperref[ass:nondegenerate-noise]{\textup{(A2)}}. 

For \(s=1,2\), Bayes' formula gives
\[
\Pr(X_1=x\mid X_2=s)
=
\frac{\Pr(X_2=0)}{\Pr(X_2=s)}
\frac{\Pr(X_2=s\mid X_1=x)}{\Pr(X_2=0\mid X_1=x)}
\Pr(X_1=x\mid X_2=0).
\]
For $s=1$ and $s=2$, the convolution formula yields
\[
\frac{\Pr(X_2=1\mid X_1=x)}{\Pr(X_2=0\mid X_1=x)}
=
\frac{e^{-\alpha x}p_1 + \alpha x e^{-\alpha x}p_0}{e^{-\alpha x}p_0}=
\frac{p_1}{p_0}+\alpha x,
\]
and
\[
\frac{\Pr(X_2=2\mid X_1=x)}{\Pr(X_2=0\mid X_1=x)}
=
\frac{p_2}{p_0}
+\alpha\frac{p_1}{p_0}x
+\frac{\alpha^2}{2}x^2,
\]
respectively.
Since \(p_0>0\), \(\alpha>0\), and \(\varepsilon_1\) is nondegenerate, \(\Pr(X_2=s)>0\) for \(s=1,2\). 
Define 
\[
c_s := \frac{\Pr(X_2=0)}{\Pr(X_2=s)}, \quad s=1, 2
\] 
and set
\[
\gamma_0 = c_1 \frac{p_1}{p_0}, \quad \gamma_1 = c_1 \alpha, \quad 
\delta_0 = c_2 \frac{p_2}{p_0}, \quad \delta_1 = c_2 \alpha \frac{p_1}{p_0}, \quad 
\delta_2 = c_2 \frac{\alpha^2}{2}. 
\]
Then, 
\(\gamma_0,\delta_0,\delta_1\ge0\) and \(\gamma_1,\delta_2>0\) 
and we have
\[
\begin{aligned}
\Pr(X_1=x\mid X_2=1)
&=
(\gamma_0+\gamma_1x)\Pr(X_1=x\mid X_2=0),\\
\Pr(X_1=x\mid X_2=2)
&=
(\delta_0+\delta_1x+\delta_2x^2)\Pr(X_1=x\mid X_2=0).
\end{aligned}
\]
Multiplying by \(e^{tx}\), summing over \(x\in\mathbb N_0\), and using termwise differentiation gives
\[
\begin{aligned}
M_1(t)&=\gamma_0M_0(t)+\gamma_1M_0'(t),\\
M_2(t)&=\delta_0M_0(t)+\delta_1M_0'(t)+\delta_2M_0''(t),  
\end{aligned}
\]
which completes the proof. \hfill\qed
\subsubsection{Proof of Lemma~\ref{lem:reverse-ptsem-layer-identity}}
\label{proof:reverse-ptsem-layer-identity}
Under the reverse representation \eqref{eq:reverse-ptsem}, conditional on \(X_2=s\), the thinning term \(\beta\pthin X_2\) is Poisson with mean \(\beta s\) and is independent of \(\varepsilon_1^*\). Hence, for \(s=0,1,2\) and \(t\) near zero, \(M_s(t)=M_{\varepsilon_1^*}(t)\exp\{s\beta(e^t-1)\}.\)
Setting \(s=0\) gives \(M_{\varepsilon_1^*}(t)=M_0(t)\), and consequently
\[
M_s(t)=M_0(t)\exp\{s\beta(e^t-1)\},
\quad s=0,1,2.
\]
\hfill\qed
\subsubsection{Proof of Lemma~\ref{lem:three-layer-incompatibility}}
\label{proof:three-layer-incompatibility}
Assume, for contradiction, that \eqref{eq:bivariate-three-layer-tilt} and \eqref{eq:false-poisson-sink-tilt} hold simultaneously. Let \(K_0(t)=\log M_0(t)\). 
Since \(M_0'(t)/M_0(t)=K_0'(t)\) and \(M_0''(t)/M_0(t)=K_0''(t)+(K_0'(t))^2\), 
we have
\[
\begin{aligned}
&\gamma_0+\gamma_1K_0'(t)
=\exp\{\beta(e^t-1)\},\\
&\delta_0+\delta_1K_0'(t)+\delta_2\{K_0''(t)+(K_0'(t))^2\}
=\exp\{2\beta(e^t-1)\}.
\end{aligned}
\]
If \(\beta = 0\), the first identity makes \(K'_0(t)\) constant, so \(K''_0(0)=0\), contradicting the nondegeneracy of the distribution with MGF \(M_0\). Thus \(\beta>0\). The first identity then yields
\[
K_0'(t)=\frac{\exp\{\beta(e^t-1)\}-\gamma_0}{\gamma_1},\quad
K_0''(t)=\frac{\beta e^t\exp\{\beta(e^t-1)\}}{\gamma_1}.
\]
Substituting these expressions into the second identity gives
\[
\delta_0+\frac{\delta_1}{\gamma_1}\{\exp\{\beta(e^t-1)\}-\gamma_0\}
+\frac{\delta_2\beta e^t\exp\{\beta(e^t-1)\}}{\gamma_1}\\
+\frac{\delta_2}{\gamma_1^2}\{\exp\{\beta(e^t-1)\}-\gamma_0\}^2
=\exp\{2\beta(e^t-1)\}.
\]
Multiplying by \(\gamma_1^2\exp\{-\beta(e^t-1)\}\), rearranging, and setting \(z=e^t\) yields
\[
F(z):=
\omega_3\exp\{\beta(z-1)\}+\omega_2z+\omega_1+\omega_0\exp\{-\beta(z-1)\}\equiv0,
\]
where \(\omega_3=\delta_2-\gamma_1^2,\quad
\omega_2=\delta_2\gamma_1\beta>0,\quad
\omega_1=\gamma_1\delta_1-2\delta_2\gamma_0,\quad
\omega_0=\gamma_1^2\delta_0-\gamma_1\delta_1\gamma_0+\delta_2\gamma_0^2.\)
Since \(F\equiv0\) in a neighborhood of \(z=1\), 
\[
F''(1)=\beta^2(\omega_3+\omega_0)=0,\quad
F'''(1)=\beta^3(\omega_3-\omega_0)=0.
\]
Hence \(\omega_3=\omega_0=0\), and then \(F(z)=\omega_2z+\omega_1\equiv0\), which implies \(\omega_2=0\), contradicting \(\omega_2=\delta_2\gamma_1\beta>0\). Therefore no such \(\beta\ge0\) exists. 
\hfill\qed
\subsubsection{Proof of Lemma~\ref{lem:local-three-layer-bayes-tilt}}
Since \(U\) is parent-closed, \(\bm X_U\) follows the PT-SEM associated with the induced subDAG \(G[U]\). 
For \(k\in U\), write \(p_k(r)=\Pr(\varepsilon_k=r)\) and \(q_k(y;\lambda)=\Pr\left(\operatorname{Poisson}(\lambda)+\varepsilon_k=y\right)\), so that
\begin{equation*}
q_k(0;\lambda)=e^{-\lambda}p_k(0),\quad
\frac{q_k(1;\lambda)}{q_k(0;\lambda)}
=\frac{p_k(1)}{p_k(0)}+\lambda,\quad
\frac{q_k(2;\lambda)}{q_k(0;\lambda)}
=\frac{p_k(2)}{p_k(0)}
+\frac{p_k(1)}{p_k(0)}\lambda
+\frac{\lambda^2}{2}.
\tag{A.1}
\label{eq:local-q-ratios}
\end{equation*}
Empty sums and products are interpreted as \(0\) and \(1\). Fix \(x\) in the support of \(p_i\), and let \(\bm x_U^{(s)}\) denote the configuration with \(x_i^{(s)}=x\), \(x_c^{(s)}=s\), and \(x_r^{(s)}=0\) for \(r\in R\).
Define
\[
\lambda_k^{(s)}
=
\sum_{j\in\operatorname{pa}_{G[U]}(k)}
\alpha_{kj}x_j^{(s)}.
\]
Since the joint distribution of the induced PT-SEM factorizes according to $G[U]$, we have
\[
\Pr(X_i=x,L_s)
=
p_i(x)q_c(s;\alpha_{ci}x)
\left\{
\prod_{r \in R} p_r(0)
\right\}
\exp\left[
-x \sum_{r \in R} \alpha_{ri} -s \sum_{r \in R} \alpha_{rc}
\right],
\tag{A.2}
\label{eq:local-joint-mass-ratio-product}
\]
\[
\frac{\Pr(X_i=x,L_s)}{\Pr(X_i=x,L_0)}
=
\exp\left\{-s\sum_{r\in R}\alpha_{rc}\right\}
\frac{q_c(s;\alpha_{ci}x)}{q_c(0;\alpha_{ci}x)},
\quad s=1,2.
\tag{A.3}
\label{eq:local-joint-mass-ratios}
\]
Substituting \eqref{eq:local-q-ratios} into \eqref{eq:local-joint-mass-ratios} yields
\[
\begin{aligned}
\frac{\Pr(X_i=x,L_1)}{\Pr(X_i=x,L_0)}
&=
\exp\left\{-\sum_{r\in R}\alpha_{rc}\right\}
\left\{
\frac{p_c(1)}{p_c(0)}+\alpha_{ci}x
\right\},\\
\frac{\Pr(X_i=x,L_2)}{\Pr(X_i=x,L_0)}
&=
\exp\left\{-2\sum_{r\in R}\alpha_{rc}\right\}
\left\{
\frac{p_c(2)}{p_c(0)}
+\alpha_{ci}\frac{p_c(1)}{p_c(0)}x
+\frac{\alpha_{ci}^2}{2}x^2
\right\}.
\end{aligned}
\tag{A.4}
\label{eq:local-explicit-joint-mass-ratios}
\]

Since $q_c(0;\alpha_{ci}x)=e^{-\alpha_{ci}x}p_c(0)$ by \eqref{eq:local-q-ratios}, 
\[
\Pr(X_i=x,L_0)
=
p_i(x)p_c(0)
\left\{\prod_{r\in R}p_r(0)\right\}
\exp\left\{
-\left(\alpha_{ci}+\sum_{r\in R}\alpha_{ri}\right)x
\right\}.
\tag{A.5}
\label{eq:local-baseline-joint-mass}
\]
In particular, 
\(\Pr(X_i=0,L_0)>0\) by the assumption \hyperref[ass:zero-mass]{\textup{(A1)}}, and hence \(\Pr(L_0)>0\). 
\eqref{eq:local-joint-mass-ratio-product} also shows that \(X_i\mid L_0\) has the same support as \(\varepsilon_i\).
The conditional distribution defining \(M_0\) is therefore nondegenerate. 

Since
\(
\alpha_{ci}+\sum_{r\in R}\alpha_{ri}\geq\alpha_{ci}>0,
\)
for sufficiently small \(|t|\) and \(j=0,1,2\), the summands defining the \(j\)-th derivative of \(M_0\) are dominated by a constant multiple of
\[
x^j
\exp\left\{
-\frac{1}{2}
\left(\alpha_{ci}+\sum_{r\in R}\alpha_{ri}\right)x
\right\}
p_i(x).
\]
The factor preceding \(p_i(x)\) is bounded on \(\mathbb N_0\), so termwise differentiation is valid and
\[
M_0^{(j)}(t)
=
\sum_{x\geq0}
x^j e^{tx}\Pr(X_i=x\mid L_0),
\quad j=0,1,2.
\tag{A.6}
\label{eq:local-mgf-derivatives}
\]

By assumptions \hyperref[ass:zero-mass]{\textup{(A1)}} and \hyperref[ass:nondegenerate-noise]{\textup{(A2)}}, there exists \(m\geq1\) such that \(p_i(m)>0\). \eqref{eq:local-baseline-joint-mass} gives \(\Pr(X_i=m,L_0)>0\), and then 
\eqref{eq:local-explicit-joint-mass-ratios} 
implies
\[
\Pr(X_i=m,L_1)>0, \quad \Pr(X_i=m,L_2)>0.
\]
Thus \(\Pr(L_1)>0\) and \(\Pr(L_2)>0\).

For \(x\) in the support of \(X_i\mid L_0\), Bayes' formula gives
\[
\Pr(X_i=x\mid L_s)
=
\frac{\Pr(L_0)}{\Pr(L_s)}
\frac{\Pr(X_i=x,L_s)}{\Pr(X_i=x,L_0)}
\Pr(X_i=x\mid L_0),
\quad s=1,2. 
\tag{A.7}
\label{eq:local-bayes-reweighting}
\]
Combining \eqref{eq:local-explicit-joint-mass-ratios} and \eqref{eq:local-bayes-reweighting}, there exist coefficients satisfying
\(
\gamma_0,\delta_0,\delta_1\geq0,
\) and \(
\gamma_1,\delta_2>0,
\)
such that
\[
\begin{aligned}
\Pr(X_i=x\mid L_1)
&=
(\gamma_0+\gamma_1x)\Pr(X_i=x\mid L_0),\\
\Pr(X_i=x\mid L_2)
&=
(\delta_0+\delta_1x+\delta_2x^2)\Pr(X_i=x\mid L_0).
\end{aligned}
\tag{A.8}
\label{eq:local-conditional-mass-identities}
\]
The strict inequalities follow from \(\alpha_{ci}>0\), while all Bayes normalizing factors are positive.
Outside this support, by \eqref{eq:local-joint-mass-ratio-product}, 
\[
\Pr(X_i=x, L_0)=\Pr(X_i=x, L_1)=\Pr(X_i=x, L_2)=0, 
\]
so \eqref{eq:local-conditional-mass-identities} remains valid for every \(x\in\mathbb N_0\). 

Multiplying \eqref{eq:local-conditional-mass-identities} by \(e^{tx}\), summing over \(x\geq0\), and using \eqref{eq:local-mgf-derivatives} gives
\[
M_1(t)=\gamma_0M_0(t)+\gamma_1M_0'(t),
\quad
M_2(t)=\delta_0M_0(t)+\delta_1M_0'(t)+\delta_2M_0''(t)
\]
for all \(t\) in a neighborhood of zero. 
\hfill\qed
\subsubsection{Proof of Proposition~\ref{prop:conditional-cgf-sink}}
Suppose first that \(i\) is a sink in \(G[U]\). Since \(U\) is parent-closed, 
\[
p_{X_i\mid\bm X_{U\setminus\{i\}}}
\left(x_i\mid\bm x\right)
=
p_{X_i\mid\bm X_{\operatorname{pa}_{G[U]}(i)}}
\left(
x_i\mid
\bm x_{\operatorname{pa}_{G[U]}(i)}
\right)
\]
for every \(\bm x\) with
\(
\Pr\!\left( \bm X_{U\setminus\{i\}}=\bm x \right)>0. 
\)
By the definition of PT-SEM, 
for all \(t\) in a neighborhood of zero,
\[
\begin{aligned}
\mathbb{E}\!\left[e^{tX_i}\mid\bm X_{U\setminus\{i\}}=\bm x\right]
&=M_{\varepsilon_i}(t)\exp\!\left\{(e^t-1)\sum_{j\in\operatorname{pa}_{G[U]}(i)}\alpha_{ij}x_j\right\},
\end{aligned}
\]
and hence 
\[
K_i^U(t\mid\bm x)=K_{\varepsilon_i}(t)+(e^t-1)\sum_{j\in\operatorname{pa}_{G[U]}(i)}\alpha_{ij}x_j.
\]
Thus \eqref{eq:sink-conditional-cgf} holds with \(g=K_{\varepsilon_i}\), \(b_j=\alpha_{ij}\) for \(j\in\operatorname{pa}_{G[U]}(i)\), and \(b_j=0\) otherwise. Assumption \hyperref[ass:nondegenerate-noise]{\textup{(A2)}} ensures that \(K_{\varepsilon_i}(t)\) is the CGF of a nondegenerate count-valued distribution.

Conversely, suppose that \(i\) is not a sink in \(G[U]\). 
By Lemma~\ref{lem:local-three-layer-bayes-tilt} 
\[
\begin{aligned}
M_1(t)
&=
\gamma_0M_0(t)+\gamma_1M_0'(t),\\
M_2(t)
&=
\delta_0M_0(t)+\delta_1M_0'(t)+\delta_2M_0''(t),
\end{aligned}
\]
for some \(\gamma_0,\delta_0,\delta_1\geq0\) and \(\gamma_1,\delta_2>0\).

If \(i\) also satisfied \eqref{eq:sink-conditional-cgf}, applying it to the conditioning values \((X_c,\bm X_R)=(s,\bm 0_R)\) defining \(L_s\), \(s=0,1,2\), would give
\[
M_s(t)
=
M_0(t)\exp\!\left\{s b_c(e^t-1)\right\},
\quad
s=0,1,2.
\]
This is exactly \eqref{eq:false-poisson-sink-tilt} with \(\beta=b_c\ge0\).
Lemma~\ref{lem:three-layer-incompatibility} continues to hold when \(M_s(t)\) is defined by
\eqref{eq:cond_mgf_sink}, yielding a contradiction.
\hfill\qed
\subsubsection{Proof of Lemma~\ref{lemma:cov_positive_definite}}

Let \(\bm a=(a_i)_{i\in S}\neq\bm 0\). Choose \(r\in S\) with \(a_r \ne 0\) such that $r=\max\{\,i\in S:a_i\neq0\,\}$. Every vertex \(i\in S\) with \(a_i\neq0\) other than \(r\) precedes
\(r\), and hence \(X_i\) is independent of \(\varepsilon_r\). Define
\[
W
=
\sum_{i\in S\setminus\{r\}}a_iX_i
+
a_r\sum_{j<r}\alpha_{rj}\pthin X_j.
\]
\(W\) is independent of \(\varepsilon_r\), and the PT-SEM for \(X_r\) gives \(\sum_{i\in S}a_iX_i=W+a_r\varepsilon_r\). Therefore,
\[
\begin{aligned}
\bm a^\top\bm\Sigma_{SS}\bm a
=
\operatorname{Var}\!\left(\sum_{i\in S}a_iX_i\right)
=
\operatorname{Var}(W)
+
a_r^2\operatorname{Var}(\varepsilon_r)
>0,
\end{aligned}
\]
where the strict inequality follows from \(a_r\neq0\) and \hyperref[ass:nondegenerate-noise]{\textup{(A2)}}. Therefore, \(\bm\Sigma_{SS}\) is positive definite.
\hfill\qed
\subsubsection{Proof of Theorem~\ref{thm:ptsem-exact-identifiability}}
 
Let \(1,\ldots,d\) be a topological order recovered by recursively applying
Proposition~\ref{prop:conditional-cgf-sink}, and set
\(S_k=\{1,\ldots,k-1\}\). 
For \(S_k \ne \varnothing\), the covariance identity \eqref{eq:ptsem-moment}, together with the independence of \(\varepsilon_k\) from its predecessors, yields, for each \(\ell\in S_k\),
\[
\Cov(X_k,X_\ell)
=
\sum_{j<k}\Cov(\alpha_{kj}\pthin X_j,X_\ell)
=
\sum_{j<k}\alpha_{kj}\Cov(X_j,X_\ell).
\]
Equivalently, \(\bm\Sigma_{S_k k}=\bm\Sigma_{S_kS_k}\bm\alpha_{k,S_k}.\) By Lemma~\ref{lemma:cov_positive_definite}, \(\bm\Sigma_{S_kS_k}\) is invertible, so \(\bm\alpha_{k,S_k}=\bm\Sigma_{S_kS_k}^{-1}\bm\Sigma_{S_k k}\) is uniquely determined by the observational distribution. Its positive entries identify the parents of \(X_k\), and their values are the corresponding thinning coefficients. 
Therefore, both the DAG and the thinning coefficient matrix are uniquely determined. 

For each \(k\), the event \(\{\bm X_{S_k}=\bm 0_{S_k}\}\) has positive probability: by induction along the recovered order, the event \(\{\varepsilon_j=0,\ j\in S_k\}\) implies \(\bm X_{S_k}=\bm 0_{S_k}\), and this event has positive probability by \hyperref[ass:zero-mass]{\textup{(A1)}} and mutual independence. Given \(\bm X_{S_k}=\bm 0_{S_k}\), all incoming thinning terms into \(X_k\) vanish, and the conditional distribution of \(X_k\) given \(\bm X_{S_k}=\bm 0_{S_k}\) coincides with the distribution of \(\varepsilon_k\). Thus, the node-wise exogenous distributions are also determined.
\hfill\qed
\subsubsection{Proof of Proposition~\ref{prop:binomial-poisson-obstruction}}
\label{app:proof-binomial-nonpoisson-identifiability}

\noindent\textit{1.}
Suppose that \(i\) is a sink in \(G[U]\). Then \(X_i\mid\left( \bm X_{U\setminus\{i\}}=\bm x\right)
\ =\
\sum_{j\in\operatorname{pa}_{G[U]}(i)}\alpha_{ij}\bthin x_j+\varepsilon_i,\) 
\[
K_i^U(t\mid\bm x)=K_{\varepsilon_i}(t)+\sum_{j\in\operatorname{pa}_{G[U]}(i)}x_j\log(1-\alpha_{ij}+\alpha_{ij}e^t).
\]
This is \eqref{eq:binomial-sink-cgf} with
\[
g(t)=K_{\varepsilon_i}(t),\quad
b_j=
\begin{cases}
\alpha_{ij}, & j\in\operatorname{pa}_{G[U]}(i),\\
0, & j\in U\setminus\bigl(\operatorname{pa}_{G[U]}(i)\cup\{i\}\bigr).
\end{cases}
\]

\noindent\textit{2.}
Suppose that \eqref{eq:binomial-sink-cgf} holds for the nonsink \(i\). 
By \hyperref[ass:zero-mass]{\textup{(A1)}}, \(\mathrm{Pr}(\bm X_{U \setminus \{i\}}=\bm 0)>0\). 
Conditional on $\bm X_{U\setminus\{i\}}=\bm 0$, we have $X_h=0$ for every $h \in \operatorname{ch}_{G[U]}(i)$. If $\alpha_{hi}=1$ for some $h \in \operatorname{ch}_{G[U]}(i)$, then $X_h=0$ implies $X_i=0$, since $1\circ_b X_i=X_i$ and all other terms in the structural equation for $X_h$ are nonnegative. Hence, the conditional distribution of $X_i$ given $\bm X_{U\setminus \{i\}}=\bm0$ is degenerate at zero, contradicting \eqref{eq:binomial-sink-cgf} with nondegenerate $g(t)$.

Choose \(c\in\operatorname{ch}_{G[U]}(i)\) such that no other child of \(i\) is a descendant of \(c\). Let \(D\) be the set of proper descendants of \(c\) in \(G[U]\), and set \(W=U\setminus D\). Then \(W\) is parent-closed, \(c\) is a sink in \(G[W]\), and no vertex in \(D\) is a parent or child of \(i\), or a parent of one of its children. For any \(j\in D\), assumptions \hyperref[ass:zero-mass]{\textup{(A1)}} and \hyperref[ass:nondegenerate-noise]{\textup{(A2)}} ensure that there exists an \(m_j >0 \) such that \(\Pr(\varepsilon_j=m_j)>0\). 
Setting $\varepsilon_j=m_j$ and $\varepsilon_k=0$ for $k\in U\setminus\{j\}$ implies $X_j=m_j$ and $X_k=0$ for $k\in U\setminus D$.
Hence, there exists a positive-probability realization 
\(\bm x^{(j)}\) of \(\bm X_{U\setminus\{i\}}\) such that
\[
x_j^{(j)}=m_j, \quad x_k^{(j)}=0,\; k\in W \setminus \{i\}.
\]
By the DAG factorization, the conditional distribution of $X_i \mid \bm{X}_{U \setminus \{i\}}$ depends on the conditioning values only through the parents of \(i\), its children, and the other parents of its children. 
Since none of these vertices belong to \(D\), and 
\(\bm x^{(j)}\) and \(\bm 0\) agree outside \(D\), we have
\[
\Pr\left(
X_i=x
\mid
\bm X_{U\setminus\{i\}}=\bm x^{(j)}
\right)
=
\Pr\left(
X_i=x
\mid
\bm X_{U\setminus\{i\}}=\bm 0
\right),
\quad x\in\mathbb N_0.
\]
Consequently, $K_i^U(t\mid\bm x^{(j)})=K_i^U(t\mid\bm 0)$. Applying \eqref{eq:binomial-sink-cgf} to both sides and using $x_\ell^{(j)}=0$ for $\ell\notin D$ yields
\[
\sum_{\ell\in D}
x_\ell^{(j)}
\log(1-b_\ell+b_\ell e^t)=0.
\]
Since \(b_\ell\in[0,1]\), all summands are nonnegative for sufficiently small \(t>0\). As \(x_j^{(j)}=m_j>0\), the identity forces \(b_j=0\), and hence \(b_\ell=0\) for every \(\ell\in D\) since \(j\in D\) was arbitrary.

For \(s=0,1\), define
\[
L_s=\{X_c=s,\ \bm X_{W\setminus\{i,c\}}=\bm 0\},\quad M_s(t)=\mathbb E[e^{tX_i}\mid L_s]. 
\]
Both $L_0$ and $L_1$ have positive probabilities. 
Writing \(p_k(r)=\Pr(\varepsilon_k=r)\), define
\[
\zeta:= \prod_{h\in\operatorname{ch}_{G[W]}(i)} (1-\alpha_{hi}) \in (0,1).
\]
Conditional on $X_i=x$, the probability that all Bernoulli offspring variables from $i$ to its children are zero is $\zeta^x$. Hence, by the DAG factorization,
\[
\Pr(X_i=x\mid L_0)
=
\frac{p_i(x)\zeta^x}
{\sum_{y\geq0}p_i(y)\zeta^y}.
\]

Under $L_1$, $X_j = 0$ for $j \in \operatorname{pa}_{G[W]}(c)\setminus\{i\}$. Hence, conditional on $X_i=x$, $X_c=1$ occurs either when $\varepsilon_c=1$ and the Bernoulli offspring variable on $i \to c$ is zero, or when $\varepsilon_c=0$ and the offspring variable equals one. Therefore,
\begin{equation}
    \tag{A.9}
    \label{eq:prob_ratio}
    \frac{\Pr(X_i=x,L_1)}{\Pr(X_i=x,L_0)}
    =   
    \frac{
    p_c(1)(1-\alpha_{ci})^x
    +
    p_c(0)x\alpha_{ci}(1-\alpha_{ci})^{x-1}
    }{  
    p_c(0)(1-\alpha_{ci})^x
    }
    =
    \frac{p_c(1)}{p_c(0)}
    +
    \frac{\alpha_{ci}}{1-\alpha_{ci}}x.
\end{equation}

From \eqref{eq:prob_ratio} and Bayes' rule, we obtain $M_1(t)=\gamma_0 M_0(t)+\gamma_1M_0'(t)$ for some \(\gamma_0\geq0\) and \(\gamma_1>0\). On the other hand, \eqref{eq:binomial-sink-cgf} gives
$M_1(t)=M_0(t)(1-b_c+ b_c e^t)$, where $b_c\in[0,1]$. Thus, with \(K_0(t)=\log M_0(t)\), these two representations imply
\[
\frac{M_1(t)}{M_0(t)}
=
\gamma_0+\gamma_1K_0'(t)
=
1-b_c+b_c e^t.
\]
The case \(b_c=0\) would make \(K_0'(t)\) constant and \(M_0\) degenerate.
Hence \(b_c>0\), and integration using \(K_0(0)=0\) gives
\[
K_0(t)=at+\lambda(e^t-1),\quad
a=\frac{1-b_c-\gamma_0}{\gamma_1},\quad
\lambda=\frac{b_c}{\gamma_1}>0.
\]
Then \(M_0(t)=e^{at}\exp\{\lambda(e^t-1)\}\), which is the MGF of \(Z+a\), where $Z\sim\operatorname{Poisson}(\lambda)$. Since \(X_i\mid L_0\) is supported on \(\mathbb N_0\) and $\Pr(X_i=0\mid L_0)>0$, we must have \(a=0\). Hence, \(M_0(t)=\exp\{\lambda(e^t-1)\}\) and $X_i\mid L_0\sim\operatorname{Poisson}(\lambda)$. Therefore,
\[
\frac{p_i(x)\zeta^x}
{\sum_{y\geq0}p_i(y)\zeta^y}
=
e^{-\lambda}\frac{\lambda^x}{x!},
\quad x\in\mathbb N_0,
\]
and hence \(p_i(x)
\propto
\frac{(\lambda/\zeta)^x}{x!}.\)
Normalizing over \(x\in\mathbb N_0\), \(p_i(x)\) is the \(\operatorname{Poisson}(\lambda/\zeta)\) pmf:
\[
p_i(x)
=
\exp\!\left(-\frac{\lambda}{\zeta}\right)
\frac{(\lambda/\zeta)^x}{x!},
\quad x\in\mathbb N_0.
\]
\hfill\qed
\subsubsection{Proof of Theorem~\ref{thm:binomial-nonpoisson-identifiability}}
By Proposition~\ref{prop:binomial-poisson-obstruction}, \eqref{eq:binomial-sink-cgf} characterizes the sinks in every parent-closed induced subDAG, since a nonsink of $G[U]$ is also a nonsink of $G$. The remainder of the proof of Theorem~\ref{thm:ptsem-exact-identifiability} applies with $\circ_p$ replaced by $\circ_b$: binomial thinning satisfies \(\operatorname{Cov}(\alpha\circ_b X,Y)=\alpha\,\operatorname{Cov}(X,Y)\), and the argument of Lemma~\ref{lemma:cov_positive_definite} still gives the invertibility of the covariance submatrices.
\hfill\qed
\subsection{Proofs for Section~\ref{sec:learning}}
\label{app:proofs-section4}

\subsubsection{Proof of Lemma~\ref{lem:moment_finiteness_ptsem}}
By \hyperref[ass:local-mgf]{\textup{(A3)}}, for every \(j\in V\), \(M_{\varepsilon_j}(t)<\infty\) for all sufficiently small \(t>0\). We proceed by induction along a topological ordering of \(G\). For a root node \(k\), the claim follows from \(X_k=\varepsilon_k\). Suppose that \(k\) is a nonroot node and that, for every \(j\in\operatorname{pa}_G(k)\), there exists \(t_j^\star>0\) such that \(M_{X_j}(t_j^\star)<\infty\). Then,
\[
\mathbb{E}\!\left[e^{tX_k}\mid\bm X_{\operatorname{pa}_G(k)}\right]
=
M_{\varepsilon_k}(t)\exp\left\{(e^t-1)\sum_{j\in\operatorname{pa}_G(k)}\alpha_{kj}X_j\right\}.
\]

Let \(d_k=|\operatorname{pa}_G(k)|\). Choose \(t>0\) sufficiently small that \(M_{\varepsilon_k}(t)<\infty\) and \(d_k\alpha_{kj}(e^t-1)\le t_j^\star\) for every \(j\in\operatorname{pa}_G(k)\). H\"older's inequality then yields
\[
\mathbb{E}\!\left[\exp\!\left((e^t-1)\sum_{j\in\operatorname{pa}_G(k)}\alpha_{kj}X_j\right)\right]
\le
\prod_{j\in\operatorname{pa}_G(k)}
\left\{
\mathbb{E}\!\left[\exp\!\left(d_k\alpha_{kj}(e^t-1)X_j\right)\right]
\right\}^{1/d_k}
<\infty.
\]
Therefore \(M_{X_k}(t)<\infty\) for some \(t>0\). Induction shows that, for every \(k\in V\), there exists \(t_k^\star>0\) such that \(\mathbb{E}[e^{t_k^\star X_k}]<\infty\). Finally, \(e^{t_k^\star x}\ge(t_k^\star x)^r/r!\) for every \(r\in\mathbb N\) and \(x\ge0\); hence,
\[
\mathbb{E}[X_k^r]
\le
\frac{r!}{(t_k^\star)^r}\mathbb{E}[e^{t_k^\star X_k}]
<\infty.
\]
\hfill\qed
\subsubsection{Proof of Proposition~\ref{prop:plugin-rate}}
For each node \(k\), write \(S_k=\operatorname{pa}_G(k)\), and let
\(\bm\alpha_{k,S_k}\) denote the corresponding population thinning coefficients. Let \(\bm\alpha_G\), \(\tilde{\bm\alpha}_G\), and
\(\hat{\bm\alpha}_G\) denote the vectors obtained by collecting \(\bm\alpha_{k,S_k}\), \(\tilde{\bm\alpha}_{k,S_k}\), and \(\hat{\bm\alpha}_{k,S_k}\), respectively, over \(k\in V\).
For \(S\neq\varnothing\), Lemma~\ref{lemma:cov_positive_definite} implies that \(\bm\Sigma_{S_k S_k}\) is positive definite. Hence, the map $(\bm\Sigma_{S_k S_k},\bm\Sigma_{S_k k}) \mapsto \bm\Sigma_{S_kS_k}^{-1}\bm\Sigma_{S_k k}$ is continuously differentiable. The joint central limit theorem for \(\hat{\bm\nu}_N\) and the delta method therefore give \(\|\tilde{\bm\alpha}_G-\bm\alpha_{G}\|=O_p(N^{-1/2})\).  Since \(\alpha_{kj}\geq 0\), the nonnegative truncation satisfies
\[
\vert\hat\alpha_{kj}-\alpha_{kj}\vert
=
\left\vert\max\left(\tilde\alpha_{kj},0\right)-\alpha_{kj}\right\vert
\leq
\vert\tilde\alpha_{kj}-\alpha_{kj}\vert.
\]
Hence, \(\bigl\|\hat{\bm\alpha}_G-\bm\alpha_G\bigr\|\leq\bigl\|\tilde{\bm\alpha}_G-\bm\alpha_G\bigr\|=O_p(N^{-1/2}).\)

By \eqref{eq:exogenous-moment-estimates},
\(\hat m_k(S_k)\) and \(\hat v_k(S_k)\) are polynomial in the empirical moments and
\(\hat{\bm\alpha}_G\).
Since both the empirical moments and \(\hat{\bm\alpha}_G\) have \(O_p(N^{-1/2})\) errors, it follows that
\[
\left|\hat m_k(S_k)-E[\varepsilon_k]\right|=O_p(N^{-1/2}),
\quad
\left|\hat v_k(S_k)-\operatorname{Var}(\varepsilon_k)\right|=O_p(N^{-1/2}).
\]
Continuous differentiability of the continuous parameter coordinates of \(H_{\tau_k}\) transfers this rate by the delta method, while local constancy of its discrete coordinates gives exact recovery with probability tending to one. Since \(d\) is fixed, stacking over \(k\in V\) yields
\(
\left\|\hat{\bm\eta}^{\,\mathrm{MoM}}_m-\bm\eta_{m,0}\right\|
=
O_p(N^{-1/2}).
\)
\hfill\qed
\subsubsection{Proof of Proposition~\ref{prop:mom-bic-gap}}

We first establish the local likelihood bounds needed below. Fix a correctly specified candidate model \(m=(G,\bm\tau)\) and a node \(k\), set \(S=\operatorname{pa}_G(k)\) and \(r=\tau_k\), and write \(\bm\alpha=\bm\alpha_{k,S}\) and \(\bm\theta=\bm\theta_k\). Define
\[
g_y(\lambda,\bm\theta)=\sum_{t=0}^y e^{-\lambda}\frac{\lambda^t}{t!}p_r(y-t;\bm\theta),\quad \lambda=\bm\alpha^\top\bm x_S.
\]
With \(g_y=0\) for \(y<0\), \(\partial_\lambda g_y=g_{y-1}-g_y\) and \(\partial_\lambda^2g_y=g_{y-2}-2g_{y-1}+g_y\). Let \(T=\sum_{j\in S}\alpha_{kj}\circ_p X_j\) be the total thinning contribution. Conditional on \(\bm X_S=\bm x_S\), \(T\sim\operatorname{Pois}(\lambda)\), \(X_k=T+\varepsilon_k\), and \(T\indep\varepsilon_k\). For fixed \(\bm x_S\), if the value of \(\lambda\) at \(\bm\eta_{m,0}\) is positive, then \(\lambda\) is bounded away from zero in a sufficiently small feasible neighborhood, and
\begin{align*}
\lambda\frac{g_{y-1}}{g_y}&=\mathbb{E}[T\mid X_k=y,\bm X_S=\bm x_S]\le y, \\
\lambda^2\frac{g_{y-2}}{g_y}&=\mathbb{E}[T(T-1)\mid X_k=y,\bm X_S=\bm x_S]\le y(y-1).
\end{align*}
If instead the value of \(\lambda\) at \(\bm\eta_{m,0}\) is zero, then, conditional on \(\bm X_S=\bm x_S\), \(T=0\) under the population parameter and hence \(X_k=\varepsilon_k\). Using the locally fixed support and the mass-ratio bounds in \hyperref[ass:bic-regularity]{\textup{(A5)}}, termwise comparison of \(g_{y-h}\) and \(g_y\) gives polynomial bounds for \(g_{y-h}/g_y\), \(h=1,2\), uniformly in \(\lambda\ge0\). Thus, in either case, these ratios are locally bounded by polynomials in the observed counts. Together with the derivative bounds in \hyperref[ass:bic-regularity]{\textup{(A5)}}, this yields polynomial envelopes for the local score and Hessian, uniformly over a fixed feasible neighborhood \(\mathcal N_m\) of \(\bm\eta_{m,0}\). Consequently, for some constants \(C<\infty\) and \(\kappa\ge0\),
\[
\sup_{\bm\eta\in\mathcal N_m}\left\{\|\nabla\log p_m(\bm X;\bm\eta)\|+\|\nabla^2\log p_m(\bm X;\bm\eta)\|\right\}\le C(1+\|\bm X\|)^\kappa, 
\]
where $\nabla$ and $\nabla^2$ denote differentiation with respect to the continuous coordinates of $\bm\eta$.

By Lemma~\ref{lem:moment_finiteness_ptsem}, the envelope on the right-hand side is square-integrable and hence integrable. Write \(Q_m(\bm\eta)=\mathbb E_0[\log p_m(\bm X;\bm\eta)]\). Differentiation under expectation is  valid, and
\[
\nabla\ell_m(\bm\eta_{m,0})-N\nabla Q_m(\bm\eta_{m,0})=O_p(N^{1/2}),\quad
\sup_{\bm\eta\in\mathcal N_m}\|\nabla^2\ell_m(\bm\eta)\|=O_p(N),
\]
where derivatives in zero-thinning coordinates are understood as right derivatives.

Let \(\bm\Delta_m=\hat{\bm\eta}^{\,\mathrm{MoM}}_m-\bm\eta_{m,0}\). Proposition~\ref{prop:plugin-rate} gives \(\|\bm\Delta_m\|=O_p(N^{-1/2})\). With probability tending to one, the discrete coordinates equal their population values, hence the expansions below involve only the continuous coordinates. Lemma~\ref{lem:moment_finiteness_ptsem} gives finite marginal means and hence finite marginal entropies, so entropy subadditivity yields \(H(P_0)<\infty\). Since \(m\) is correctly specified,
\[
Q_m(\bm\eta)-Q_m(\bm\eta_{m,0})=-D_{\mathrm{KL}}\!\left(P_0\,\|\,P_{m,\bm\eta}\right)\le0,\quad \bm\eta\in\Theta_m.
\]
Hence every interior continuous component of \(\nabla Q_m(\bm\eta_{m,0})\) is zero, while for any redundant edge \(j\to k\) with population coefficient \(\alpha_{kj}=0\), the feasible right derivative with respect to \(\alpha_{kj}\) is nonpositive. Since the nonnegative truncation gives \(\hat\alpha_{kj}\ge0\), \(\nabla Q_m(\bm\eta_{m,0})^\top\bm\Delta_m\le0. \)

As \(\|\Delta_m\|=O_p(N^{-1/2})\), the segment joining \(\eta_{m,0}\) and \(\hat{\eta}^{\mathrm{MoM}}_m\) lies in \(\mathcal N_m\) with probability tending to one. A second-order expansion therefore gives
\begin{align*}
\ell_m\!\left(\hat{\bm\eta}^{\,\mathrm{MoM}}_m\right)-\ell_m(\bm\eta_{m,0})&=N\nabla Q_m(\bm\eta_{m,0})^\top\bm\Delta_m+O_p(N^{1/2})\|\bm\Delta_m\|+O_p(N)\|\bm\Delta_m\|^2
\\
&=N\nabla Q_m(\bm\eta_{m,0})^\top\bm\Delta_m+O_p(1).
\end{align*}
Since \(\nabla Q_m(\bm\eta_{m,0})^\top\bm\Delta_m\le0\), the positive part of this difference is \(O_p(1)\). As \(m\) is correctly specified, \(\ell_m(\bm\eta_{m,0})=\ell_0\), proving the second assertion. For \(m_0\), all edge coefficients are positive and all continuous family coordinates are interior, so \(\nabla Q_{m_0}(\bm\eta_0)=0\). Therefore, 
\(
\ell_{m_0}\!\left(\hat{\bm\eta}^{\,\mathrm{MoM}}_{m_0}\right)=\ell_0+O_p(1),
\)
which proves the first assertion. 
\hfill\qed

\subsubsection{Proof of Theorem~\ref{thm:bic-consistency}}

First, let \(m=(G,\bm\tau)\) be a correctly specified candidate model. If \(G=G_0\) and \(\bm\tau\in\mathcal T_0\), no comparison is needed; hence suppose \(G\neq G_0\) or \(\bm\tau\notin\mathcal T_0\).
By removing the zero-coefficient edges, we obtain a PT-SEM representation of \(P_0\), whose support graph must be \(G_0\) by Theorem~\ref{thm:ptsem-exact-identifiability}.
Thus, \(G\) contains \(G_0\), with all additional edges having population coefficient zero, so that \(|E|\ge |E_0|\), with strict inequality if \(G\neq G_0\).
Moreover, the exogenous distributions in any correctly specified representation are the true node-wise exogenous distributions, and by the definition of \(\mathcal T_0\), \(\sum_{k\in V}q_{\tau_k}\ge\sum_{k\in V}q_{\tau_{0,k}}\), with strict inequality if \(\bm\tau\notin\mathcal T_0\). Hence,
\[
q(m)
=|E|+\sum_{k\in V}q_{\tau_k}
\ge q(m_0)+1.
\]
Proposition~\ref{prop:mom-bic-gap} therefore gives
\[
\operatorname{BIC}^{\mathrm{MoM}}(m)-\operatorname{BIC}^{\mathrm{MoM}}(m_0)\ge \log N-O_p(1)\xrightarrow{p}+\infty.
\]

Next, let \(m\) be misspecified, and write 
\(
Q_m(\bm\eta)=\mathbb E_0[\log p_m(\bm X;\bm\eta)], \ Q_0=\mathbb E_0[\log p_{m_0}(\bm X;\bm\eta_0)]. 
\)
The proof of Proposition~\ref{prop:mom-bic-gap} gives \(H(P_0)<\infty\), and hence \(Q_0=-H(P_0)>-\infty\).

For any candidate edge \(j\to k\), \(X_j\ge\varepsilon_j\) and \hyperref[ass:nondegenerate-noise]{\textup{(A2)}} imply
\(\Pr_0(X_j\ge1)>0\).
Hence, there exists \(B_{kj}\in\mathbb N_0\) such that
\(\pi_{kj}:=\Pr_0(A_{kj})>0\), where \(A_{kj}:=\{X_j\ge1,\ X_k\le B_{kj}\}\).
Write
$\lambda_k=\bm\alpha_{k,\operatorname{pa}_G(k)}^\top \bm X_{\operatorname{pa}_G(k)}$. 
On \(A_{kj}\), \(\lambda_k\ge\alpha_{kj}\), and for some finite \(C_{kj}\),
\begin{equation}
\label{ineq:log_pm}
p_{k,\tau_k}\!\left(
X_k\mid\bm X_{\operatorname{pa}_G(k)};
\bm\alpha_{k,\operatorname{pa}_G(k)},\bm\theta_k
\right)
\le
e^{-\lambda_k}\sum_{t=0}^{B_{kj}}\frac{\lambda_k^t}{t!}
\le
C_{kj}e^{-\alpha_{kj}/2}.
\tag{A.10}
\end{equation}
Since every node-wise log-pmf is nonpositive,
\begin{align*}
\log p_m(\bm X;\bm\eta)
&=
\sum_{\ell\in V}
\log p_{\ell,\tau_\ell}
\!\left(
X_\ell\mid
\bm X_{\operatorname{pa}_G(\ell)};
\bm\alpha_{\ell,\operatorname{pa}_G(\ell)},
\bm\theta_\ell
\right)\\
&\le
\log p_{k,\tau_k}
\!\left(
X_k\mid
\bm X_{\operatorname{pa}_G(k)};
\bm\alpha_{k,\operatorname{pa}_G(k)},
\bm\theta_k
\right).
\end{align*}
Hence,
\[
\begin{aligned}
Q_m(\bm\eta)
&\le
E_0\!\left[
\log p_{k,\tau_k}
\!\left(
X_k\mid
\bm X_{\operatorname{pa}_G(k)};
\bm\alpha_{k,\operatorname{pa}_G(k)},
\bm\theta_k
\right)
\right]
\\
&\le
E_0\!\left[
\bm 1_{A_{kj}}
\log p_{k,\tau_k}
\!\left(
X_k\mid
\bm X_{\operatorname{pa}_G(k)};
\bm\alpha_{k,\operatorname{pa}_G(k)},
\bm\theta_k
\right)
\right]
\\
&\le
\pi_{kj}
\left(
\log C_{kj}
-\frac{\alpha_{kj}}{2}
\right), 
\end{aligned}
\]
so \(Q_m(\bm\eta)\to-\infty\) whenever any thinning coefficient diverges. 
Therefore, when \(\sup_{\bm\eta \in \Theta_m} Q_m > -\infty\), the thinning coefficients may be restricted to a bounded set. Together with \hyperref[ass:bic-regularity]{\textup{(A5)}}, this allows the supremum of \(Q_m\) to be taken over a compact subset of \(\Theta_m\).

By \hyperref[ass:bic-regularity]{\textup{(A5)}}, for each fixed value of the discrete parameters and fixed $\bm x$, $p_m(\bm x;\bm\eta)$ is continuous in the continuous parameters. Hence, $\log p_m(\bm x;\bm\eta)$ is upper semicontinuous in the continuous parameters, with $\log 0=-\infty$. Thus, for any sequence $\bm\eta_n\to\bm\eta$ with the discrete parameters fixed,
\[
-\log p_m(\bm X;\bm\eta)
\le
\liminf_{n\to\infty}
\{-\log p_m(\bm X;\bm\eta_n)\}
\quad P_0\text{-a.s.}
\]
Since $-\log p_m\ge0$, Fatou's lemma gives
$-Q_m(\bm\eta) \le \liminf_{n\to\infty}\{-Q_m(\bm\eta_n)\}$, or equivalently, $Q_m(\bm\eta) \ge \limsup_{n\to\infty}Q_m(\bm\eta_n)$. 
Since the discrete parameter components take values in finite sets, if
$\sup_{\bm\eta\in\Theta_m} Q_m(\bm\eta)>-\infty$, the supremum is attained at some \(\bm\eta_m^*\) in the compact subset considered above.

Since $m$ is misspecified,
$P_{m,\bm\eta_m^*}\ne P_0$, we have
\[
\sup_{\bm\eta\in\Theta_m} Q_m(\bm\eta)
=
Q_m(\bm\eta_m^*)
=
Q_0-D_{\mathrm{KL}}(P_0\Vert P_{m,\bm\eta_m^*})
<Q_0.
\]
If $\sup_{\bm\eta\in\Theta_m} Q_m(\bm\eta)=-\infty$, the above inequality is immediate because $Q_0>-\infty$. Hence, in either case, there exists $\delta_m>0$ such that
\[
\sup_{\bm\eta\in\Theta_m}Q_m(\bm\eta)
\le Q_0-\delta_m.
\]


We next establish a corresponding bound for the sample log-likelihood. 
Let \(\bm X^{(1)},\ldots,\bm X^{(N)}\) be an i.i.d.\ sample from \(P_0\), and define
\(
N_{kj}:=\sum_{i=1}^N \mathbf 1\{\bm X^{(i)}\in A_{kj}\}.
\)
Then, by the law of large numbers, $N_{kj}/N\to\pi_{kj}>0$ in probability.
Using \eqref{ineq:log_pm} for these $N_{kj}$ observations and the fact that all log-pmf terms are nonpositive, we obtain
\[
\frac{1}{N}\ell_m(\bm\eta)
\le
\frac{N_{kj}}{N}
\left(
\log C_{kj}-\frac{\alpha_{kj}}{2}
\right).
\]
Hence, for each candidate edge $j\to k$, we can choose a finite constant $R_{kj}>0$ sufficiently large so that, with probability tending to one,
\[
\alpha_{kj}>R_{kj}
\;\Rightarrow\;
\frac{1}{N}\ell_m(\bm\eta) \le Q_0-\frac{\delta_m}{2}.
\]

Since there are only finitely many candidate edges, define
\[
\mathcal K_m
=
\left\{
\bm\eta=(\bm\alpha,\bm\theta)\in\Theta_m:
0\le\alpha_{kj}\le R_{kj}, 
\text{ for every }j\to k\in E
\right\}.
\]
By the compactness of the parameter spaces of the candidate exogenous families in \hyperref[ass:bic-regularity]{\textup{(A5)}}, \(\mathcal K_m\) is compact. Moreover,
\begin{equation}
\label{eq:misspecified-outside-compact}
\Pr\left(
\sup_{\bm\eta\in\Theta_m\setminus\mathcal K_m}
\frac{1}{N}\ell_m(\bm\eta)
\le Q_0-\frac{\delta_m}{2}
\right)
\to 1.
\tag{A.11}
\end{equation}
It remains to control the sample log-likelihood uniformly over $\mathcal K_m$. 
For $M>0$, define
\[
f_{\bm\eta,M}(\bm x) := \max\{\log p_m(\bm x;\bm\eta),-M\} \ge \log p_m(\bm x;\bm\eta). 
\]
Fix $\bm\eta_0\in\mathcal K_m$. Since
\[
f_{\bm\eta_0,M}(\bm x)
\downarrow
\log p_m(\bm x;\bm\eta_0), 
\quad \text{as }M\to\infty,
\]
the monotone convergence theorem gives
\[
E_0[f_{\bm\eta_0,M}(\bm X)]
\downarrow
Q_m(\bm\eta_0)
\le Q_0-\delta_m.
\]
We may therefore choose $M=M(\bm\eta_0)$ sufficiently large so that
\[
E_0[f_{\bm\eta_0,M}(\bm X)]
<
Q_0-\frac{3\delta_m}{4}.
\]

For this fixed $M$, let
\(
\overline U_r(\bm\eta_0)
:=
\left\{
\bm\eta\in\mathcal K_m:
\|\bm\eta-\bm\eta_0\| \le r
\right\},
\)
where, if discrete parameter components are present, they are fixed at their values in $\bm\eta_0$.
By upper semicontinuity of $f_{\bm\eta,M}(\bm x)$ in the continuous parameter components,
\[
\sup_{\bm\eta\in\overline U_r(\bm\eta_0)}
f_{\bm\eta,M}(\bm x)
\;\rightarrow\;
f_{\bm\eta_0,M}(\bm x)
\quad \text{as}\quad r\downarrow0.
\]
Since $-M \le f_{\bm\eta,M} \le 0$, the bounded convergence theorem yields
\[
E_0\left[
\sup_{\bm\eta\in\overline U_r(\bm\eta_0)}
f_{\bm\eta,M}(\bm X)
\right]
\;\rightarrow\;
E_0[f_{\bm\eta_0,M}(\bm X)]
\quad\text{as }r\downarrow0.
\]
Thus, we can choose $r>0$ sufficiently small and set $U_{\bm\eta_0}:=\{\bm\eta\in\mathcal K_m:\|\bm\eta-\bm\eta_0\|\ < r\}$, with the discrete parameter components fixed as above, so that
\[
E_0\left[
\sup_{\bm\eta\in U_{\bm\eta_0}}
f_{\bm\eta,M}(\bm X)
\right]
<
Q_0-\frac{\delta_m}{2}.
\]

The neighborhoods $U_{\bm\eta_0}$, $\bm\eta_0\in\mathcal K_m$, form an open cover of $\mathcal K_m$. By compactness, there exist $\bm\eta_1,\ldots,\bm\eta_L \in \mathcal{K}_m$ such that
$
\mathcal K_m
\subseteq
\bigcup_{\ell=1}^L U_{\bm\eta_\ell}.
$
Let $M_\ell=M(\bm\eta_\ell)$, and define
\[
g_\ell(\bm x)
=
\sup_{\bm\eta\in U_{\bm\eta_\ell}}
f_{\bm\eta,M_\ell}(\bm x),
\quad
\ell=1,\ldots,L.
\]
By construction,
\[
E_0[g_\ell(\bm X)]
<
Q_0-\frac{\delta_m}{2},
\quad
\ell=1,\ldots,L.
\]
Moreover, since \(-M_\ell\le g_\ell(\bm x)\le0\), each \(g_\ell\) is bounded.

For any $\bm\eta\in\mathcal K_m$, there exists some $\ell$ such that
$\bm\eta\in U_{\bm\eta_\ell}$, and hence
\[
\log p_m(\bm x;\bm\eta)
\le
f_{\bm\eta,M_\ell}(\bm x)
\le
g_\ell(\bm x).
\]

Therefore, since each $g_\ell$ is bounded and $L$ is finite, the law of large numbers gives
\[
\sup_{\bm\eta\in\mathcal K_m}\frac{1}{N}\ell_m(\bm\eta)
\le
\max_{1\le\ell\le L}\frac{1}{N}\sum_{i=1}^N g_\ell(\bm X^{(i)})
=
\max_{1\le\ell\le L}E_0[g_\ell(\bm X)]+o_p(1)
\le
Q_0-\frac{\delta_m}{2}+o_p(1).
\]

Combining this with \eqref{eq:misspecified-outside-compact}, and since
\(\hat{\bm\eta}^{\,\mathrm{MoM}}_m\in\Theta_m\) by construction,
\[
\ell_m\!\left(\hat{\bm\eta}^{\,\mathrm{MoM}}_m\right)\le N\!\left(Q_0-\frac{\delta_m}{2}\right)+o_p(N).
\]
Proposition~\ref{prop:mom-bic-gap} and the law of large numbers give \(\ell_{m_0}(\hat{\bm\eta}^{\,\mathrm{MoM}}_{m_0})=NQ_0+o_p(N)\). Therefore,
\[
\operatorname{BIC}^{\mathrm{MoM}}(m)-\operatorname{BIC}^{\mathrm{MoM}}(m_0) 
\ge 
N \delta_m +o_p(N) 
\xrightarrow{p}+\infty.
\]

Since \(d\) and \(\mathcal M\) are fixed, there are finitely many candidate models. Thus, with probability tending to one, every candidate with \(G\neq G_0\) or \(\bm\tau\notin\mathcal T_0\) has larger \(\operatorname{BIC}^{\mathrm{MoM}}\) than \(m_0\). Therefore, \(\Pr(\hat G=G_0,\hat{\bm\tau}\in\mathcal T_0)\to1\).
\hfill\qed
\subsubsection{Computational complexity}
\label{app:complexity}

Let \(d=|V|\) and let \(N\) be the sample size. We measure complexity in \(d\) and \(N\), treating the fixed candidate family collection and a single local conditional-pmf evaluation as constant-cost. There are \(d2^{d-1}\) node--parent-set pairs. For a parent set of size \(s\), computing the MoM estimate and local score costs \(O(s^3+Ns+N)\); hence all local scores can be computed in \(O\!\left((d^4+Nd^2)2^d\right)\) time. The subset-DP recursions require \(O(d^22^d)\) time and are therefore of lower order. Thus the overall time complexity is \(O\!\left((d^4+Nd^2)2^d\right)\). The local-score and DP tables require \(O(d2^d)\) storage, while retaining the coefficient vectors for all node--parent-set pairs requires \(O(d^22^d)\). Hence, the overall space complexity is \(O(d^22^d)\).
\bibliography{PTSEM}
\end{document}